\documentclass[11pt]{article}
\usepackage{amsmath}
\usepackage{graphicx}
\usepackage{bm}

\usepackage[dvips]{color}
\usepackage{amssymb}
\usepackage{amsfonts}
\usepackage{comment}

\def\be{\begin{equation}}
\def\ee{\end{equation}}
\def\ba{\begin{eqnarray}}
\def\ea{\end{eqnarray}}

\title{Field Equation of Correlation Function of Mass Density Fluctuation for Self-Gravitating Systems }
\author{\small   Yang  Zhang\thanks{yzh@ustc.edu.cn}
                  \, and Qing  Chen\thanks{cqpb@mail.ustc.edu.cn}           \\
       \small Key Laboratory for Researches in Galaxies and Cosmology, \\
     \small Department of  Astronomy,  University of Science and Technology of China, \\
     \small Hefei, Anhui, 230026,  China }

\def\be{\begin{equation}}
\def\ee{\end{equation}}
\def\ba{\begin{eqnarray}}
\def\ea{\end{eqnarray}}

\sf

\begin{document}

\maketitle

\begin{abstract}
We study the mass density distribution of
the Newtonian self-gravitating system.
Modeling the system either as a gas in thermal equilibrium,
or as a fluid in hydrostatical equilibrium,
we obtain the field equation of  correlation function $\xi(r)$ of
the mass density fluctuation itself.
It can apply to the study of galaxy clustering on Universe large scales.
The observed $\xi(r)\simeq (r_0/r)^{1.7}$ follows from first principle.

The equation tells that
$\xi(r)$ depends on the point mass $m$ and Jeans wavelength scale $\lambda_{0}$,
which are different for galaxies and clusters.
It explains several longstanding, prominent features
of the observed clustering:
the profile of $\xi_{cc}(r)$ of clusters
is similar to $\xi_{gg}(r)$  of galaxies
but with a higher amplitude and a longer correlation length,
 the correlation length increases with the mean separation between
clusters  $r_0\simeq 0.4d$ as the observed scaling,
and on very large scales $\xi_{cc}(r)$  exhibits periodic oscillations
with a characteristic wavelength $\sim 120$Mpc.
With a set of fixed  model parameters,
the solution $\xi(r)$ for galaxies and for clusters,
the power spectrum,
the projected, and the angular correlation function,
simultaneously agree with the observational data
from the surveys, such as Automatic Plate Measuring (APM),
Two-degree-Field Galaxy Redshift Survey (2dFGRS), and
Sloan Digital Sky Survey (SDSS), etc.
\end{abstract}

\section{Introduction}

To understand the  matter distribution in Universe on large scales is
one of the major goals of modern cosmology.
The large scale structure is determined by self gravity of galaxies and clusters.
It brings interest to the study of self-gravitating systems.
Since the number of galaxies as the typical objects  is enormous,
one needs statistics to study the distribution.
In this regard,
the $2$-point correlation function $\xi_{gg}(r)$ of galaxies
and $\xi_{cc}(r)$ of clusters
serve as a powerful statistical tool \cite{Bok,Peebles}.
It  not only provides the statistical information,
but also contains the underlying dynamics mainly due to gravitational force.
Therefore, we would like to investigate the correlation functions of
self gravitating system in thermal equilibrium for the first step although the real
Universe is not in thermal equilibrium.

Over the years,
various observational surveys have been carried out
for galaxies and for clusters,
such as the Automatic Plate Measuring (APM) galaxy survey
                   \cite{loveday1996stromlo},
the Two-degree-Field Galaxy Redshift Survey (2dFGRS)\cite{peacock2001measurement},
Sloan Digital Sky Survey (SDSS)\cite{abazajian2009seventh}, etc.
All these surveys suggest that the correlation of galaxies
has a  power law form $\xi_{gg}(r)\propto (r_0/r)^{\gamma}$ with $r_0\sim 5.4h^{-1}$Mpc
and $\gamma\sim 1.7$ in a range $ (0.1\sim 10)h^{-1}$Mpc
\cite{TotsujiKihara,groth1977large,Peebles,GrothPeebles1986,SoneraPeebles}.
The correlation of clusters is found to be
of a similar form:  $\xi_{cc}(r)\sim 20 \xi_{gg}(r)$
 in a range $ (5\sim 60)h^{-1}$Mpc,
with an amplified magnitude \cite{1983ApJ...270...20B,KlypinKopylov}.
For quasars $\xi_{qq}(r)\sim 5 \xi_{gg}(r)$\cite{Shaver}.

In the past,
numerical computations and simulations
have been extensively employed to study the clustering of galaxies and of clusters,
and significant progresses have been made.
To understand the physical mechanism behind the clustering,
 analytical studies are important.
In particular,
Reference \cite{Saslaw68,Saslaw69,Saslaw85,Saslaw} used macroscopic thermodynamic variables,
such as internal energy, entropy, pressure, etc,
for  adequate descriptions,
whereby
the power-law form of $\xi_{gg}(r)$
was introduced as modifications to the energy and pressure.
Similarly, Reference \cite{deVegaSanchez,deVegaSanchez98}  used
the grand partition function of the self-gravitating gas
to study a possible fractal structure of
the correlation function of galaxies.
However the field equation of $\xi$
was not given in these studies.

Also adopting statistical mechanics,
we employ the techniques of the generating functional $Z[J]$.
This practice has been well known in
particle physics and condensed matter physics.
The key point is that we express $Z[J]$ as
a path integral over the mass density field $\psi$,
instead of the gravitational potential.
The functional derivatives of $\ln Z[J]$
give the connected Green functions
$G^{(n)}(r_1,...,r_n)=\langle \delta\psi(r_1)...\delta\psi(r_n) \rangle $,
i.e., the correlation functions of the density fluctuation
$\delta\psi$ about the mean density $\psi_0=\langle \psi  \rangle $
\cite{zhang,zhang2009nonlinear}.
In order to set up the field equation of the $2$-pt correlation function $G^{(2)}(r)$,
we first derive the field equation of the mass density field $\psi$,
which is equivalent to the well-known Lane-Emden equation
for the gravitational field  \cite{Emden}.
The use of the density field $\psi$ suits our purpose.
This has been achieved by modeling the system either as a gas in thermal equilibrium,
or as a fluid at rest in the gravitational field in hydrostatical equilibrium.
The equation of $\psi$ is highly nonlinear.
To deal with this issue, we apply the perturbation method,
let $\psi=\psi_0+\delta\psi$,
and expand the equation in terms of small quantity $\delta\psi$.
We keep only up to $(\delta \psi) ^2$ and drop off higher order terms.
By taking the ensemble average of the field equation
of  $\psi(\bf r)$,
and taking functional derivative $\delta/\delta J(\bf r)$ of the averaged equation,
the {\it field equation} of  $G^{(2)}(r)$
is derived.
The  advantage of this formulation is that
the field equation of $G^{(n)}(r_1,...,r_n)$ for any $n$
can be also derived systematically.
As is anticipated,
the 3-point correlation function $G^{(3)}$ also appear
in the field equation of $G^{(2)}$ to this order of perturbations.
To cut off the hierarchy,
$G^{(3)}$ can be expressed as the products of $G^{(2)}$
by the Kirkwood-Groth-Peebles ansatz  \cite{Kirkwood,groth1977large}.
In the procedure,
the quantities like  $G^{(2)}(0)$,  $\nabla G^{(2)}(0)$, and $\nabla^2 G^{(2)}(0)$
also show up, as always happens for any interacting field theory
when going to high orders of perturbations.
After necessary renormalization to absorb these quantities,
we end up with the nonlinear field equation of $G^{(2)}(r)$, also denoted as $\xi(r)$,
with three parameters, $a$, $b$, $c$,
as the coefficients of nonlinear terms beyond the Gaussian approximation.

The formulation applies to the system of galaxies and
to the system of clusters as well,
whereby
the particle mass $m$ and the Jeans wavelength $\lambda_J$
can vary  in the field equation.
With a  set of fixed values of $a$, $b$, $c$,
the solution $\xi(r)$ will confront  simultaneously
the observational data of galaxies and of clusters.
For galaxies,
this will also be done for the power spectrum,
the projected, and angular correlation functions.
This work surpasses the previous sketched work  \cite{zhang,zhang2009nonlinear}
by presenting the detailed derivation of the field equation,
the renormalization,  and modifications of new nonlinear terms.
Besides, this work also presents
the projected, and angular correlation functions,
and their direct comparisons with the observations.

In section 2, we shall derive
the field equation of $\psi(\bf r)$ by hydrostatics,
and write down the generating functional $Z[J]$.

In section 3, we  shall  derive the nonlinear field equation of $\xi(r)$.

In section 4, by inspecting the resulting  equation of $\xi(r)$,
we shall give its several predictions
about the prominent features of galaxy correlation, cluster correlation,
and the large scale structure.

In section 5, we  shall  present the solution $\xi(r)$
for a fixed set of parameters $(a,b,c)$,
and confront with the observed correlation function for galaxies.
Similar comparisons will be carried out
to the power spectrum, the projected, and angular correlation functions,
correspondingly.

In section 6 we  shall  apply the same solution $\xi(r)$
 with a greater mass $m$
to the system of clusters,
and compare with the observational data of clusters.
The observed scaling of the ``correlation length" $r_0$ will be explained
and the observed $\sim 120$Mpc periodic oscillations will be interpreted.

Section 7 contains  conclusions, discusses.

In Appendix A, we give the formulation of the grand partition function
of the self-gravitating system in terms of path integral
over the gravitational field.

In Appendix B,  by the technique of functional differentiation,
we present the comprehensive details of the derivation
of the field equation of $G^{(2)}(r)$
and its renormalization involved.

We use a unit with the speed of light $c=1$
and the Boltzmann constant $k_B=1$.

\section{Field Equation of Mass Density of Self-Gravitating System }

Galaxies, or clusters, distributed in Universe
can be approximately described
as a fluid at rest in the gravitational field due to the fluid,
i.e, by hydrostatics.
This modeling is  an approximation
since the cosmic expansion is not considered.
As has been discussed by Saslaw \cite{Saslaw},
the system of galaxies in the expanding Universe
is in an asymptotically relaxed state, i.e, a quasi thermal equilibrium,
since the cosmic time scale
$1/H_0$ is longer than the local crossing time scale.
Therefore, the hydrostatic approximation is appropriate
for a preliminary study of this paper.

In general,
a fluid is described by the continuity equation, the Euler equation,
and the Poisson equation :
\be
\frac{\partial \rho}{\partial t} +\nabla \cdot (\rho {\bf v})=0,
\ee
\be
\frac{\partial \bf v}{\partial t}+({\bf v}\cdot \nabla){\bf v}
                =-\frac{1}{\rho}\nabla p + \nabla \Phi,
\ee
\be \label{poisson}
\nabla^2 \Phi =-4\pi G \rho.
\ee
For the hydrostatical case, $\dot\rho=0$ and ${\bf v}=0$,
the Euler equation takes the form \cite{LandauLifshitz}
\be \label{Eulereq}
\frac{1}{\rho}\nabla p = \nabla \Phi,
\ee
which describes the mechanical equilibrium of the fluid.
Denoting $c_s^2 \equiv \partial p/\partial \rho$
with $c_s$  being a constant sound speed,
Eq.(\ref{Eulereq}) becomes
\be \label{Eulereq2}
\frac{1}{\rho}\nabla \rho = \frac{1}{c_s^2}\nabla \Phi.
\ee
Taking gradient on both sides of this equation leads to
\be
\nabla^2\rho = \frac{1}{c_s^2} (\nabla\rho\cdot \nabla\Phi +\rho \nabla^2\Phi ).
\ee
Substituting Eq.(\ref{poisson}) and (\ref{Eulereq2}) into the above
gives
\be  \label{rederivation}
\nabla^{2}\rho-\frac{1}{\rho}(\nabla\rho)^{2}+\frac{4\pi G}{c_s^2}\rho^2=0.
\ee
We call Eq.(\ref{rederivation}) the field equation of  mass density
for the self-gravitating many-body system.
For convenience, we introduce a dimensionless density field
$\psi ({\bf r}) \equiv \rho({\bf r})/\rho_{0}$,
where  $\rho_0=mn_0$ is the mean mass density of the system.
Then Eq.(\ref{rederivation}) takes the form
\be \label{masseq}
\nabla^{2}\psi-\frac{1}{\psi}(\nabla\psi)^{2}+k_{J}^{2}\psi^{2}=0,
\ee
with $k_{J}\equiv \sqrt{4\pi G\rho_{0}}/c_{s}$ being the Jeans wavenumber.
This is highly nonlinear in  $\psi$ as it contains $1/\psi$.
Eq.(\ref{masseq}) also follows from $\delta\mathcal{H}(\psi)/\delta \psi=0$
with the effective Hamiltonian density
\be \label{Hpsi}
\mathcal{H}(\psi)=\frac{1}{2}(\frac{\nabla\psi}{\psi})^{2}-k_{J}^{2}\psi.
\ee
To employ
Schwinger's technique of functional derivatives  \cite{Schwinger},
we introduce an external source $J(\bf r)$ coupled to
the  field $\psi$:
\be  \label{eff_L}
\mathcal{H}(\psi,J)=\frac{1}{2}(\frac{\nabla\psi}{\psi})^{2}
               -k_{J}^{2}\psi-J\psi,
\ee
and the mass density field equation
 in the presence of  $J$ is
\be  \label{field_eq}
\nabla^{2}\psi-\frac{1}{\psi}(\nabla\psi)^{2}+k_{J}^{2}\psi^{2}+J\psi^{2}=0.
\ee
This is the key equation
we shall use in Section 3 to derive the field equation of  correlation $G^{(2)}(r)$.
The generating functional for
the correlation functions of $\psi$ is defined as
\be \label{ZJ}
Z[J] = \int  D\psi
  e^{-\alpha \int d^{3}\textbf{r}\mathcal{H}(\psi,J)},
\ee
where $\alpha  \equiv c_s^2/4\pi G m$ with $c_s $ being the sound speed
and $m$ being the mass of a single particle.
Here $\alpha$ is introduced for proper dimension.
The surveys of galaxies or clusters reveal the mass distribution,
instead of the gravitational field.
(We do not address a possible bias of mass distribution in this paper.)
The advantage of working with the mass density field $\psi$
is to confront the observational data directly \cite{zhang,zhang2009nonlinear}.

Eq.(\ref{masseq}) can also be derived from another approach.
The Universe filled with galaxies and  clusters
can be modeled as a self gravitating gas
assumed to be in thermal quasi-equilibrium \cite{Saslaw}.
Note that the Universe is expanding with a time scale  $\sim 1/H_0 =(3/8\pi G \rho_0)^{1/2}$,
and the time scale of propagation of fluctuations
 $\sim \lambda_J/c_s\sim 1/(4\pi G \rho_0)^{1/2}$,
both being of the same order of magnitude.
The thermal equilibrium is an approximation.
For such a system of $N$ particles of mass $m$,
the Hamiltonian is
\be \label{Hamiltonian}
H=\sum_{i=1}^N \frac{p_i^2}{2m}-\sum_{i<j}^N \frac{Gm^2}{r_{ij}}
\ee
with $r_{ij}=|{\bf r}_i- {\bf r}_j|$,
and the grand partition function
is
\be \label{Z1}
Z=\sum_{N=0}^\infty \frac{z^N}{N!}
\int \prod_{i=1}^N  \frac{d^3p_i\, d^3r_i}{(2\pi)^3}
                       e^{-H/T},
\ee
where $z$ is the fugacity.
Using the Stratonovich-Hubbard transformation \cite{Stratonovich,Hubbard},
$Z$ can be converted into a path integral over a field $\phi$
           \cite{deVegaSanchez,Zinn-Justin} as follows
(the detailed derivation is given in Appendix A):
\be \label{Z2}
Z=\int D\phi e^{-\alpha\int d^3 r \mathcal{H}(\phi)},
\ee
where
the effective Hamiltonian density
for $\phi$  is
\be \label{L}
\mathcal{H}(\phi)= \frac{1}{2 }(\nabla \phi)^2-k_J^2e^{\phi}.
\ee
By $\delta\mathcal{H}(\phi)/\delta \phi=0$,
Eq.(\ref{L}) yields the well-known Lane-Emden equation
\cite{Emden,Ebert,Bonnor,Antonov62,Lynden-Bell}
\be \label{Lane-Emden}
\nabla^2\phi + k^2_J e^{\phi}=0,
\ee
which, by rescaling $\phi \equiv \Phi/ c_s^2$,
becomes the Poisson equation
\be\label{Poissoneq}
\nabla^2\Phi = - 4\pi G \rho(\bf r),
\ee
where the mass density
$\rho( {\bf r} )=\rho_0 e^{\Phi({\bf r})/c_s^2}$.
Writing
\be
\psi( {\bf r})\equiv e^{\phi( {\bf r})},
\ee
Eq.(\ref{L}) and Eq.(\ref{Lane-Emden})
 become  Eq.(\ref{Hpsi}) and Eq.(\ref{masseq}), respectively,
as long as $\psi \ne 0$, i.e, $\phi\ne - \infty$.

Thus, for the self-gravitating system,
the assumption of either thermal equilibrium, or hydrostatical equilibrium,
lead to the field equation (\ref{masseq}) of mass density,
which is equivalent to the Lane-Emden equation (\ref{Lane-Emden}).
Nevertheless,
Eq.(\ref{masseq}) has the advantage that
the density field $\psi$ suits better for studying the mass distribution.

\section{Field Equation of the 2-pt Correlation Function
of Density Fluctuations}

In the following we outline the field equation of 2-pt correlation function,
and the comprehensive details are attached in Appendix B.
Since the distribution of galaxies, or clusters,
can be viewed as the fluctuations of the mass density in the homogeneous Universe,
we consider the fluctuation field
$\delta\psi(\bf{r}) \equiv  \psi(\textbf{r})-\langle\psi(\textbf{r})\rangle$,
where the statistical ensemble  average is defined as

\begin{align}
\langle\psi(\textbf{r})\rangle &=
\frac{1}{Z}\int D\psi \,\psi e^{-\alpha \int d^{3}\textbf{r}\mathcal{H}(\psi)}\nonumber\\
          &=\frac{\delta}{\alpha \delta J({\bf r })}\log Z[J] \mid_{J=0}.
\end{align}

\begin{figure}
\includegraphics[width=\linewidth]{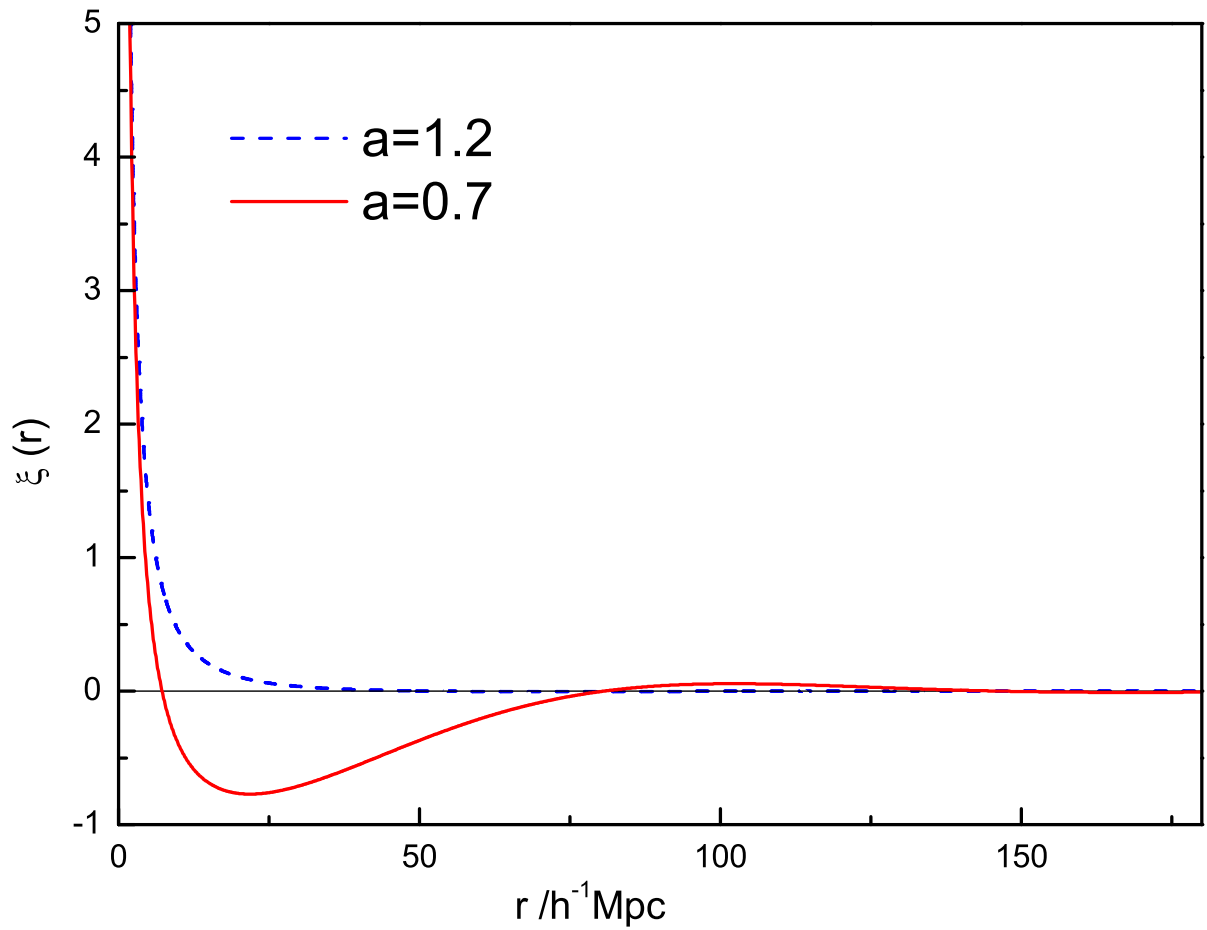}
\caption{
A large viscosity coefficient $a$
will cause strong damping to the oscillations of $\xi(r)$
at large distances.
In this graph,  $b$, $c$, and $k_0$
are fixed for demonstration.
}
\label{a}
\end{figure}

Here the subscript $|_{J=0}$ means  setting $J=0$
 after taking functional derivative.
$\langle\psi(\textbf{r})\rangle $
represents the mean of scaled mass density of the background,
and, in our case, is a constant $\langle\psi(\textbf{r})\rangle=\psi_{0}$.
The 2-point correlation function of $\delta \psi$,
i.e, the {\it connected} 2-point  Green function,
is given by the functional derivative of $\ln Z[J]$
with respect to $J$ \cite{Binney}  :

\begin{align}
G^{(2)}({\bf r}_{1},{\bf r}_{2})
&\equiv \langle\delta\psi({\bf r}_{1})\delta\psi({\bf r}_{2}) \rangle\nonumber\\
&=\alpha^{-2}
 \frac{\delta^2}{\delta J({\bf r}_1)\delta J({\bf r}_2 )}
             \log Z[J]|_{J=0}\nonumber\\
&=\alpha ^{-1}
\frac{ \delta \langle\psi({\bf r}_{2})\rangle_J }{\delta J({\bf r}_{1})}|_{J=0},
\label{2ptGreen}
\end{align}
where
$\langle\psi(\textbf{r})\rangle_J
\equiv \frac{\delta}{\alpha \delta J({\bf r })}\log Z[J]
$
before setting $J=0$.
One can take
$G^{(2)}({\bf r}_{1},{\bf r}_{2}) = G^{(2)}(r_{12}) $
for a homogeneous and isotropic Universe.
Analogously, the n-point correlation function of $\delta \psi$ is

\begin{align}
G^{(n)}({\bf r}_{1},..., {\bf r}_{n})
&\equiv \langle\delta\psi({\bf r}_{1})...\delta\psi({\bf r}_{n}) \rangle\nonumber\\
&=\alpha^{-n}
 \frac{\delta^n  \log Z[J]}{\delta J({\bf r}_1)...\delta J({\bf r}_n )}
            |_{J=0}\nonumber\\
& =\alpha^{-(n-1)}
 \frac{\delta^{n-1} \langle\psi({\bf r}_n)\rangle_J}{\delta J({\bf r}_1)
       ...\delta J({\bf r}_{n-1} )}   |_{J=0}
\end{align}

for $n\ge 3$.
To derive the field equation of $G^{(2)}(r)$, as a routine \cite{Goldenfeld},
one takes functional derivative
of the ensemble average of Eq.(\ref{field_eq})
with respect to $J({\bf r}_1)$,
\begin{align} \label{Eqintermediate}
&\frac{\delta}{\delta J({\bf r}_1)}
(\langle\nabla^{2}\psi({\bf r})\rangle_J
-\langle \frac{1}{\psi({\bf r})}(\nabla\psi({\bf r}))^{2} \rangle_J\nonumber\\
&+k_{J}^{2} \langle \psi({\bf r})^{2}\rangle_J
+J({\bf r})\langle \psi({\bf r})^{2}\rangle_J)=0,
\end{align}
and then sets $J=0$.
The detailed calculation is provided in Appendix B.
To deal with the nonlinearity of Eq.(\ref{field_eq}) systematically,
we expand it in terms of the fluctuation $\delta\psi$,
and keep up to the second order  $(\delta \psi)^2$.
Then Eq.(\ref{Eqintermediate}) leads the following equation of $G^{(2)}$:
\begin{align}   \label{2pt3pt}
&\nabla^{2}G^{(2)}(\textbf{r})+k_{0}^{2}\psi_{0}G^{(2)}(\textbf{r})\nonumber\\
&+[\frac{1}{2\psi_{0}^{2}}\nabla^{2}G^{(2)}(0)G^{(2)}(\textbf{r})-
(\frac{1}{2\psi_{0}}\nabla^{2}+k_{J}^{2})G^{(3)}(0,{\bf r},{\bf r})\nonumber\\
&+\frac{2}{\psi_{0} ^{2}}\nabla G^{(2)}(0)\cdot\nabla G^{(2)}(\textbf{r})]
=-\frac{1}{\alpha }[\psi_{0}^{2}-G^{(2)}(0)]\delta^{(3)}(\textbf{r}),
\end{align}
where the characteristic wavenumber  $k_0\equiv \sqrt{2}k_J$.
This equation is of the same form as
Eq.(4) in our previous paper \cite{zhang2009nonlinear},
except that the coefficient
of $G^{(3)}$ now acquires the $-k_{J}^{2}$ term,
and the coefficient of the source $\delta^{(3)}(\textbf{r})$
acquires   $\frac{1}{\alpha }G^{(2)}(0)$.
These modifications come from
an improved treatment to include high order contributions properly.
Note that $G^{(3)}$ occurs in Eq.(\ref{2pt3pt}).
There are various ways to cut off this  hierarchy.
In this paper,
we adopt the Kirkwood-Groth-Peebles  ansatz \cite{Kirkwood,groth1977large}
\begin{align} \label{Ansatz}
G^{(3)}(\textbf{r}_1,\textbf{r}_2,\textbf{r}_3)
&=Q[G^{(2)}(r_{12} ) G^{(2)}(r_{23} )
 +G^{(2)}(r_{23} ) G^{(2)}(r_{31} )\nonumber\\
 &+G^{(2)}(r_{31} ) G^{(2)}(r_{12} )],
\end{align}
where $Q$ is  a dimensionless parameter.
This  ansatz has been well-known and often used in studies of cosmology.
There have been abundant data from observations and simulations as well,
showing that the ansatz serves as a good fitting to the data when $Q\sim 1$.
Here we take this ansatz
because it gives a cutoff and has the connection to practice of cosmology.
Substituting Eq.(\ref{Ansatz}) into Eq.(\ref{2pt3pt}),
after a necessary renormalization to absorb
the quantities like  $G^{(2)}(0)$,  $\nabla G^{(2)}(0)$, and $\nabla^2 G^{(2)}(0)$,
we obtain the field equation of the 2-point correlation function
\begin{align} \label{eqfinal}
&(1-b\xi)\nabla^{2}\xi
+k_{0}^{2}(1- c\xi)\xi
+({\bf a}-b\nabla \xi) \cdot\nabla \xi
    =-\frac{1}{\alpha }  \delta^{(3)}(\textbf{r}),
\end{align}
where
$\xi=\xi(r) \equiv  G^{(2)}(\textbf{r})$,
and  $\bf a$, $b$, and $c$ are three independent parameters.
The special case of ${\bf a}=b=c=0$  is the Gaussian approximation,
and Eq.(\ref{eqfinal}) reduces to
 the Helmholtz equation (\ref{Helmoltz}).
Thus, the terms containing $\bf a$, $b$, and $c$
represent the nonlinear contributions beyond
the Gaussian approximation.
Eq.(\ref{eqfinal}) in the radial direction is
\be \label{final}
(1-b\xi)\xi''+
( (1-b\xi)\frac{2}{x}+a )\xi' +\xi  -b \xi'\, ^{2}   -c\xi^{2}=
-\frac{1}{\alpha }   \frac{\delta(x)k_0}{x^{2}},
\ee
where
$\xi'\equiv \frac{d}{dx}\xi$ and $x \equiv k_{0}r$.
The nonlinear terms with $b$ and $c$ in Eq.(\ref{final})
can enhance the amplitude of $\xi$ at small scales
and increase the correlation length.
The term containing $a$ plays the role of effective viscosity,
and a greater $a$ leads
a strong damping to the oscillations of $\xi$ at large scales,
as shown in Fig.\ref{a}.
The solution $\xi(r)$ will confront
the observational data of galaxies and clusters
in the following.

\section{General Predictions of  Field Equation}

Before studying its solution,
we inspect the field equation (\ref{eqfinal}) to
see its predictions about
the general properties of the correlation function $\xi(r)$.

1, The  equation contains the point mass $m$
and the characteristic wavenumber $k_0 $.
It applies to the system of galaxies,
as well as to the system of clusters,
with different respective $m$ and $k_0$ in each case.
Thus, as  solutions of Eq.(\ref{eqfinal}),
$\xi_{gg}$ for galaxies
should have a profile similar to $\xi_{cc}$  for  clusters,
but will differ in amplitude and in scale
determined by different $m$ and $k_0$.
Indeed, the observations tell that both
 $\xi_{gg}$   and $\xi_{cc}$
have a power-law form: $\propto r^{-1.8}$ in their respective, finite range,
but  $\xi_{cc}$ has a higher amplitude\cite{1983ApJ...270...20B,KlypinKopylov}.

2,  The $\delta^{3}(\bf r)$ source in Eq.(\ref{eqfinal})
has the coefficient $ 1/\alpha =4\pi G m /c_s^2$,
which determines the overall amplitude of a solution $\xi$.
The mass $m$ of a cluster can be $10 \sim 10^3$ times that of a galaxy\cite{Bahcall1999}.
As for the sound speed,
 $c_s$ can be regarded as the the peculiar velocity,
which is the same order of magnitude
 for galaxies and clusters, around several hundreds km/s
\cite{hawkins20032df,2006ApJ...653..861M}.
Therefore, $1/\alpha$ is essentially determined by $m$,
and a greater $m$ will yield a higher amplitude of $\xi$.
This property is clearer in the Gaussian approximation,
where
\be \label{xim}
\xi(r) \propto m
\ee
 as revealed by the analytical solution $\xi(r)$ seen in Eq.(\ref{GGaussian}).
This general prediction
naturally explains a whole chain of prominent facts of observations:
luminous galaxies are more massive and have a higher correlation amplitude
         than ordinary galaxies \cite{Zehavi05},
clusters are much more massive and have a much higher correlation than galaxies,
and rich clusters have a higher correlation
         than poor clusters
         since the richness $\propto $ the mass
\cite{1983ApJ...270...20B,Einasto02,Einasto2000b,Bachall03}.
         This phenomenon has been a puzzle for long \cite{Bahcall1999}
         and was interpreted as being caused by the statistics
         of rare peak events \cite{Kaiser1984}.

3, The power spectrum,
as the Fourier transform of $\xi(r)$,
is proportional to the inverse of the spatial number density:
\be \label{pkn}
P(k) \propto 1/n_0.
\ee
See in the analytical $P(k)$ in Eq.(\ref{Pkgaussian}) in the Gaussian approximation.
In fact,
given the mean mass density $\rho_0=m n_0$,
a greater $m$ implies a lower $n_0$.
Therefore, the properties  (\ref{pkn}) and (\ref{xim})
reflect the same physical law of clustering from different perspectives.
The property  (\ref{pkn}) also agrees with the observational fact
from a variety of surveys.
The observed $P(k)$ of clusters is much higher than that of galaxies,
and the observed $P(k)$ of rich clusters is higher than poor clusters, etc.
This is explained by Eq.(\ref{pkn}),
since $n_0$ of clusters is much lower than that of galaxies,
and $n_0$ of rich clusters is lower than that of poor clusters
\cite{Bahcall96,Bahcall1999}.

4,   The characteristic length $\lambda_0=2\pi/k_0
=(\frac{\pi}{2})^{1/2}\frac{c_s}{\sqrt{G\rho_0}}\\ \propto \frac{c_s}{\sqrt{\rho_0}}$
appears in Eq.(\ref{eqfinal}) as the only scale,
which underlies the scale-related features of the solution $\xi(r)$.
At a fixed $\lambda_0$, the solution $\xi(r)$ with a high amplitude
drops to its first zero at a larger distance,
leading to an apparently  longer ``correlation length".
If surveys could cover the whole Universe
and
if all the cosmic mass were in galaxies,
which, in turn,  were all contained in clusters,
then
$\rho_0$ would be the same for the system of galaxies and for the system of clusters.
Nevertheless, actual cluster surveys extend over larger spatial volumes,
including those very dilute regions.
Therefore,  $\rho_{0c}$ of
the region covered by cluster surveys
can be lower than  $\rho_{0g}$ for galaxy surveys,
and $\lambda_0 $ for cluster surveys will be longer than that for galaxy surveys,
whereas $c_s$ is roughly the same order of magnitude for galaxies and clusters.
For instance, for rich clusters,
the spatial number density $n_c\sim 10^{-5}$ clusters Mpc$^{-3}$
 compared with $n_g\sim 10^{-2} $ galaxies Mpc$^{-3}$ for
 bright galaxies, lower by three orders \cite{Bahcall96}.
But a rich cluster contains only $30\sim 300$ galaxies,
the observed mass-to-light ratio
of clusters flattens at $200\sim 300$ of the solar ratio $M/L$ \cite{BahcallLubin},
implying that clusters do not contain a substantial amount
of additional dark matter, other than that associated with the galaxy halos
and the hot intercluster medium \cite{Bahcall96}.
These imply that $\rho_{0c}$ is lower than $\rho_{0g}$.
Indeed, as will be seen in the next Section 5 and 6,
to use one solution $\xi(r)$
to match the data of both galaxies and clusters,
one has to take $k_0$ to be smaller for clusters,
than for galaxies,
so the system of clusters covered by the surveys
has a longer $\lambda_0$ than the system of galaxies
\cite{2000MNRAS.319..939C,1983ApJ...270...20B}.

\section{The Solution Confronting the Observed Data of Galaxy Surveys}

Now we give the solution $\xi_{gg}(r)$ for a fixed set of parameters
$(a,b,c)$,
and confront with the observed correlation from major galaxy surveys.
We will also convert $\xi_{gg}(r)$ into its associated power spectrum $P(k)$,
the projected correlation function $w_p(r_p)$,
and the angular correlation function $w(\theta)$,
and compare with the respective observational data, simultaneously.

1, The Correlation Function $\xi_{gg}(r)$.

\begin{figure}
\includegraphics[width=\linewidth]{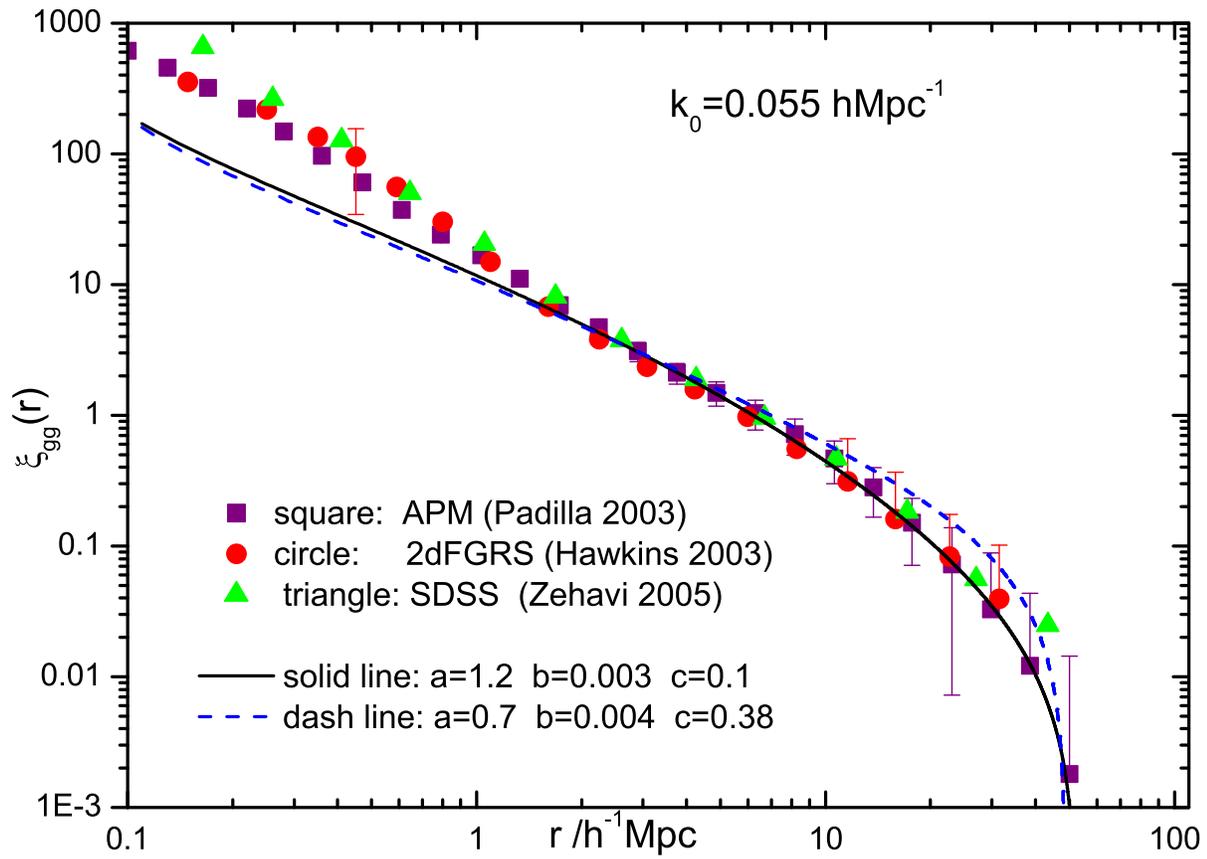}
\caption{
The solution $\xi_{gg}(r)$ confronts the data of galaxies by
 APM \cite{Padilla03},
 2dFGRS \cite{hawkins20032df}, and SDSS \cite{Zehavi05}.
 Here $k_0=0.055$ hMpc$^{-1}$ is taken in calculation.
}
\label{correlation}
\end{figure}

We have taken $k_0=0.055$hMpc$^{-1}$ for the case of galaxies.
For demonstration,
two  respective sets of the parameters are taken:
$(a,b,c)=(1.2,0.003, 0.1)$, and
$(a,b,c)=(0.7,0.004,0.38)$.
We remark that other values of $(a,b,c)$
can be also chosen to match the data.
Figure \ref{correlation} shows
the solution $\xi_{gg}(r)$ and the observed data by the galaxy surveys
of APM \cite{Padilla03}, SDSS \cite{Zehavi05}, and 2dFGRS \cite{hawkins20032df}.
It is seen that
the theoretical $\xi_{gg}(r)$
matches the observational data
 on the range of $r=(1 \sim 50)$ h$^{-1}$Mpc.
The usual power law fitting $\xi_{gg} \propto r^{-1.7}$ is valid only
 in an interval  $(0.1 \sim 10)$ h$^{-1}$Mpc.
On large scales, the solution $\xi_{gg}(r)$ deviates from the power law,
decreases rapidly to zero and becomes negative around $\sim 50$ h$^{-1}$Mpc.
However, on small scales $r \leq 1$ h$^{-1}$Mpc,
 the solution  $\xi_{gg}(r)$ is lower than the data,
even though it has already improved the Gaussian approximation\cite{zhang}.
This insufficiency at $r \leq 1$ h$^{-1}$Mpc
should be due to neglect of the high order nonlinear terms, like $(\delta \psi)^3$,
 in our perturbation.
These terms should contribute more correlations on small scales.
Notice that the scale $\sim 1$ h$^{-1}$Mpc is the size of a typical cluster,
and the high amplitude of the observed $\xi_{gg}$ at $r\leq 1$ h$^{-1}$Mpc
may come partially from
the local structure of virialized clusters.

2, The Power Spectrum $P(k)$.

\begin{figure}
\includegraphics[width=\linewidth]{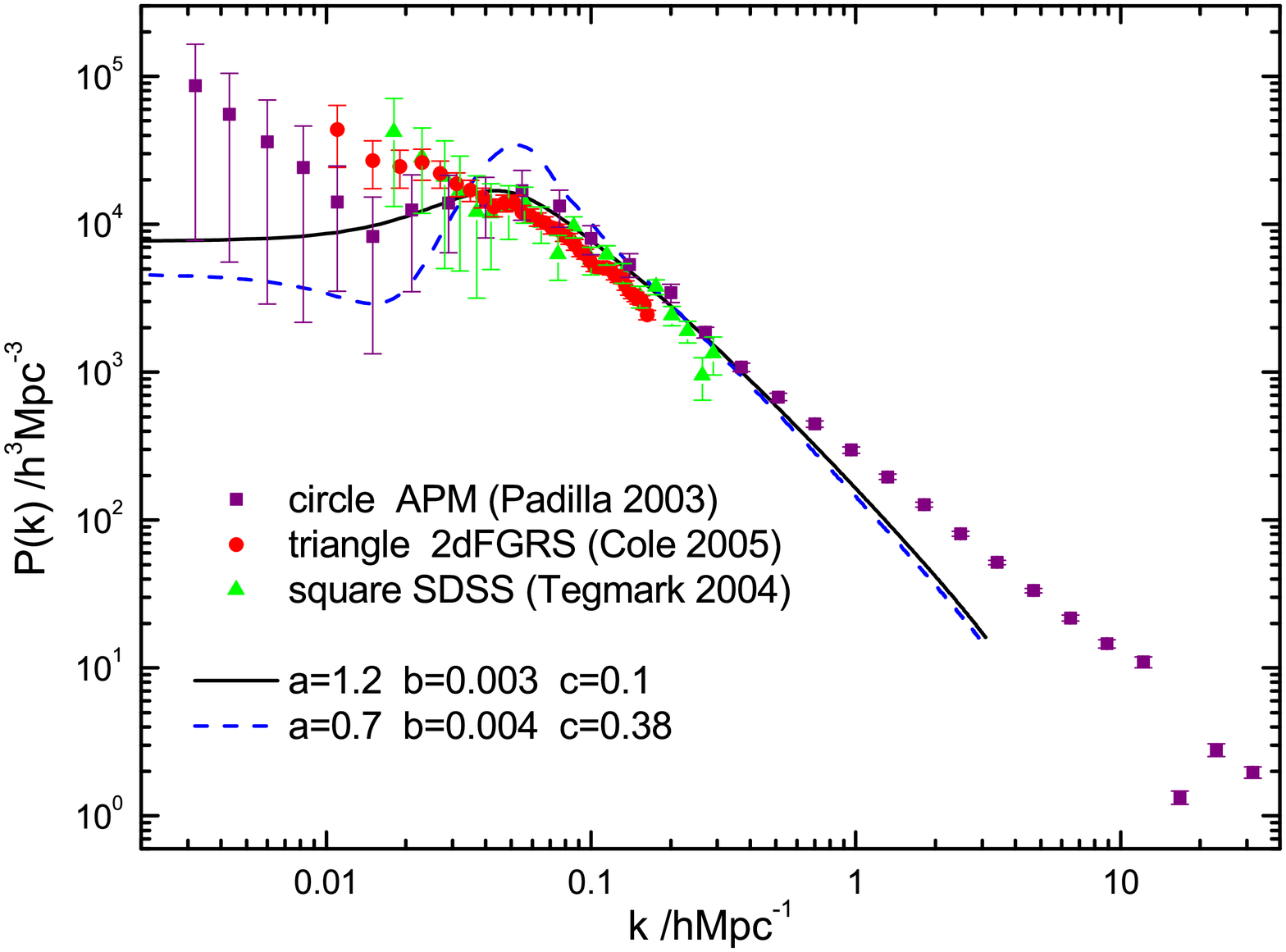}
\caption{
The power spectra $P(k)$ converted
from  $\xi_{gg}(r)$ in Figure \ref{correlation}
confronts the data of  APM \cite{Padilla03}, 2dFGRS \cite{cole20052df}
and SDSS \cite{blantonTegmark2004}.
}
\label{Pk}
\end{figure}

The power spectrum $P(k)$ is the Fourier transform
\be \label{P(k)}
P(k)=4\pi\int\limits_{0}^{\infty}\xi(r)\frac{\sin(kr)}{kr}r^{2}dr
\ee
of the  correlation function $\xi(r)$,
measuring the matter density fluctuation in the $k$-space.
In principle, $P(k)$ and $\xi(r)$ contain the same information
if both are complete on their respective space,
$k=(0, \infty)$, and $r=(0, \infty)$.
Actually,
the observed $\xi_{gg}(r)$ is not complete,
and is actually limited to a finite range, say $r \leq 50$ Mpc.
If the observed power-law $\xi_{gg}(r)=(r_0/r)^{1.8}$ were plugged in Eq.(\ref{P(k)}),
one would have $P(k)\propto k^{-1.2}$,
which does not comply with the observed $P(k)\propto k^{-1.6}$ \cite{peacockbook}.
Our solution $\xi_{gg}(r)$ is  given on the whole range $r=(0,\infty)$,
so it will yield a reliable  $P(k)$.
Figure \ref{Pk} shows the theoretical $P(k)$ converted by Eq.(\ref{P(k)})
from the solution  $\xi_{gg}(r)$
with the same set $(a,b,c)$ and $k_0$ as those in Fig.\ref{correlation}.
Also shown are the observational data of $P(k)$
from APM \cite{Padilla03}, 2dFGRS \cite{cole20052df},
and SDSS \cite{blantonTegmark2004}.
It is seen that the theoretical $P(k)$ agrees well with the data $P(k)\propto k^{-1.6}$
 in the range of $k=(0.04\sim 0.7)$ hMpc$^{-1}$.
However, at large $k$,
the theoretical $P(k)$ is lower than the data.
This insufficiency of $P(k)$ corresponds that of $\xi_{gg}(r)$
at small scales $r \leq 1$Mpc shown in Figure \ref{correlation} .
If high order terms like $(\delta\psi)^3$ are included,
the theoretical $P(k)$ is expected to improve at large $k$.

3, The Projected Correlation Function $W_p(r_p)$.

\begin{figure}
\includegraphics[width=\linewidth]{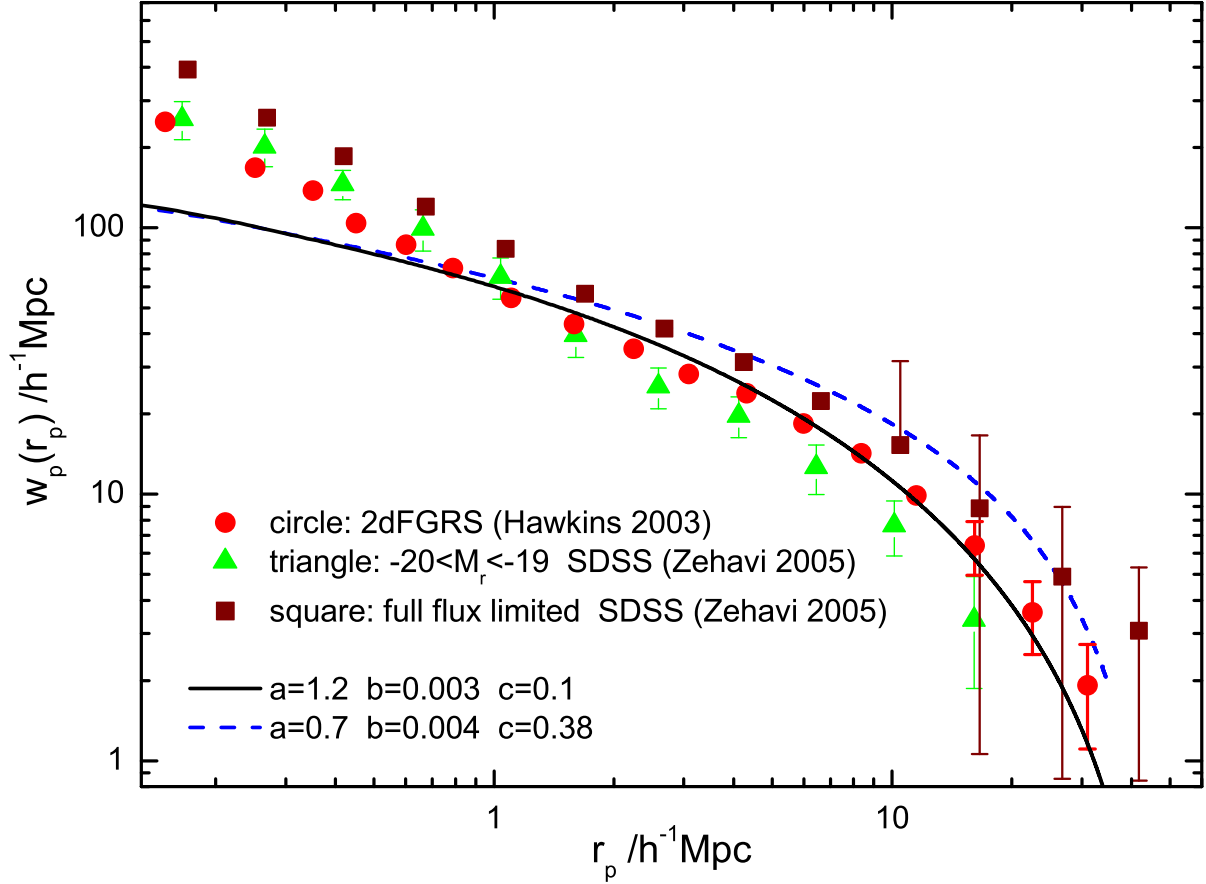}
\caption{
The projected correlation function $W_{p}(r_p)$
converted from  $\xi_{gg}(r)$ confronts
the data of
2dFGRS \cite{hawkins20032df} and SDSS \cite{Zehavi05}.
}
\label{Projection}
\end{figure}

For actual sky surveys of galaxies and clusters,
the measurement of distances is through their cosmic red-shift $z$.
The galaxies or clusters have  peculiar velocities,
causing the red-shift distortion to the measured distance.
To eliminate this distorting effect,
one can make use of the unaffected part of the correlation function
by integrating over the distance parallel to the line of sight.
This leads to the projected correlation function \cite{Peebles1976,Peebles}
\begin{align} \label{real_proj}
W_{p}(r_{p})&=2\int\limits_{0}^{\infty}\xi(\sqrt{r_{p}^{2}+y^{2}})dy
=2\int\limits_{r_{p}}^{\infty}\xi(r)\frac{rdr}{\sqrt{r^{2}-r_{p}^{2}}} \, ,
\end{align}
where $r_{p}$ is the separation of two points
vertical to the line of sight,  not distorted by the peculiar velocities.
Figure \ref{Projection} shows the theoretical $W_p(r_p)$
from the solution $\xi_{gg}$ with the same $(a,b,c)$, and $k_0$
as those in Fig.\ref{correlation}.
The observational data from
2dFGRS \cite{hawkins20032df} and SDSS \cite{Zehavi05}
are also plotted for comparison.
Overall,
the theoretical $W_{p}(r_{p})$ traces the observational data well
in the range $r_p=(0.6 \sim 30)$h$^{-1}$Mpc,
but,  is lower than the data
on small scales $r_p\leq 0.6$h$^{-1}$Mpc,
the same insufficiency  mentioned before.

4, The Angular Correlation Function $w(\theta)$.

\begin{figure}
\includegraphics[width=\linewidth]{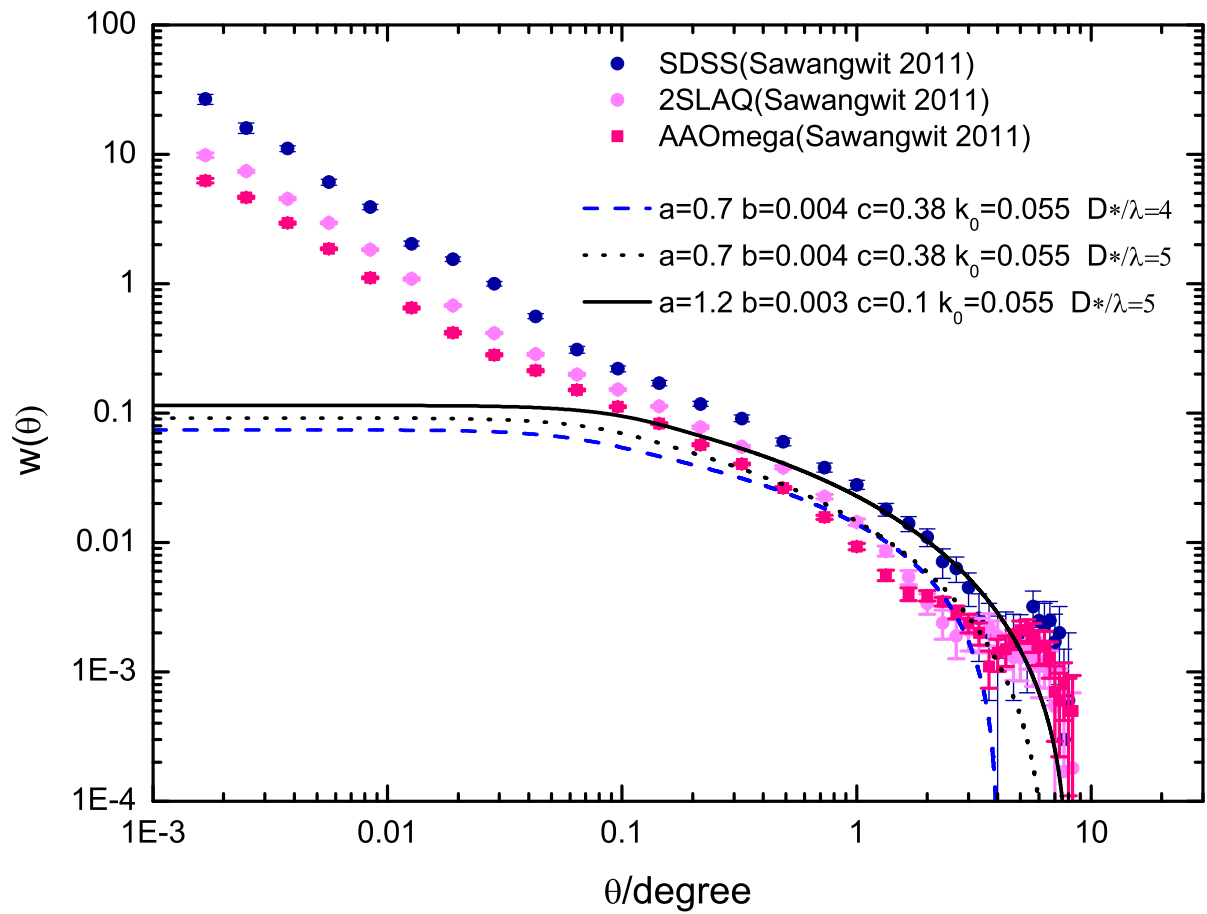}
\caption{The angular correlation function $w(\theta)$
converted from  $\xi_{gg}(r)$
confronts the data of SDSS, AA$\Omega$, 2SLAQ \cite{Sawangwit11}.}
\label{Angular}
\end{figure}

To avoid the uncertainty of the distance measurements,
similar to the projected $W_p(r_p)$,
the 2-point angular correlation function $w(\theta)$ is also used
to represent the correlation between two  angle positions.
It also involves an integration  of $\xi(r)$ along the line of sight.
Specifically, fixing the azimuth angle and leaving only the altitude $\theta$,
under the small separation approximation,
the angular correlation function $w(\theta)$ can be derived
from  $\xi(r)$ by Limber's equation \cite{Limber,Rubin,Peebles}
\be
w(\theta)=\int\limits_{0}^{\infty}y^{4}\phi(y)^{2}dy
\int\limits_{-\infty}^{\infty}\xi(\sqrt{x^{2}+\theta^{2}y^{2}})dx,
\label{angfunc}
\ee
where $\phi$ is the selection function,
representing  the combined effect of luminosity function and observer function.
With the normalization $\int\limits_{0}^{\infty}\phi y^{2}dy=1$,
it is given by \cite{peacockbook}
\be
\phi(y)=\frac{2}{\Gamma(\frac{5}{4})}
D_{\ast}^{-\frac{5}{2}}y^{-\frac{1}{2}}
e^{-(\frac{y}{D_{\ast}})^{2}},
\ee
where $D_{\ast}$ is the characteristic sample depth.
In practice, $w(\theta)$ is given by the following integration over
the wavenumber $k$ \cite{peacockbook}
\be
\resizebox{.9\hsize}{!}{$w(\theta)=\frac{\pi}{2\Gamma^{2}(\frac{5}{4})D_{\ast}}
\int\limits_{0}^{\infty}\Delta^{2}(k)\frac{dk}{k^{2}}(1-\frac{(kD_{\ast}\theta)^{2}}{8})
\exp(-\frac{(kD_{\ast}\theta)^{2}}{8}),$}
\label{w_Pk}
\ee
where $\Delta^{2}(k) \equiv k^{3}P(k)/2\pi^{2}$ and $P(k)$ is the power spectrum.
Fig. \ref{Angular} shows the calculated $w(\theta)$ by Eq.(\ref{w_Pk})
with $P(k)$ from Fig. \ref{Pk}.
It is seen that the theoretical curves trace the observed data  well
for $\theta=(0.1\sim 8)$ degree.
Also, the theoretical curve is lower than the data points
for $\theta \leq 0.1$ degree.
For a correlation length $\lambda$,
the ratio  $D_{\ast}/\lambda$ measures that
how much farther the survey goes beyond the correlated scale.
We take $\lambda=\pi/k_{0}$ for concreteness.
The survey depth of AA$\Omega$ is larger than that of SDSS \cite{Sawangwit11}.
Indeed, as shown in Fig. \ref{Angular},
to fit the data,   a larger $D_{\ast}/\lambda$  for AA$\Omega$
is required than that for SDSS.
So far, with the fixed $(a,b,c)$ and $k_0$,
the solution $\xi_{gg}(r)$
and the associated $P(k)$, $W_p(r_p)$, and $w(\theta)$
simultaneously agree with the data,  respectively,
except the insufficiency at the small scales.

\section{ Confronting the Observed Data of Clusters}

For galaxies discussed above,
the observed correlation function
is limited to $r\leq 50$h$^{-1}$Mpc.
Clusters are believed to trace the cosmic mass distribution
on even larger scales,
and the observational data cover spatial scales farther than that of galaxies.
Now we are going to apply
the solution with the same two sets of  $(a,b,c)$ as in Section 5
to the system of clusters,
each being regarded as a point mass.
The mass $m$ of a cluster is greater than that of a galaxy.
This leads to a higher overall amplitude of $\xi_{cc}(r)$,
i.e, a higher value of  the boundary condition $\xi_{cc}(r_b)$ at some point $r=r_b$.
Besides, to match the observational data of clusters,
 a small value $k_0=0.03$ Mpc$^{-1}$ is required,
smaller than the previous $k_0=0.055$Mpc$^{-1}$ for galaxies.
\begin{figure}
\includegraphics[width=\linewidth]{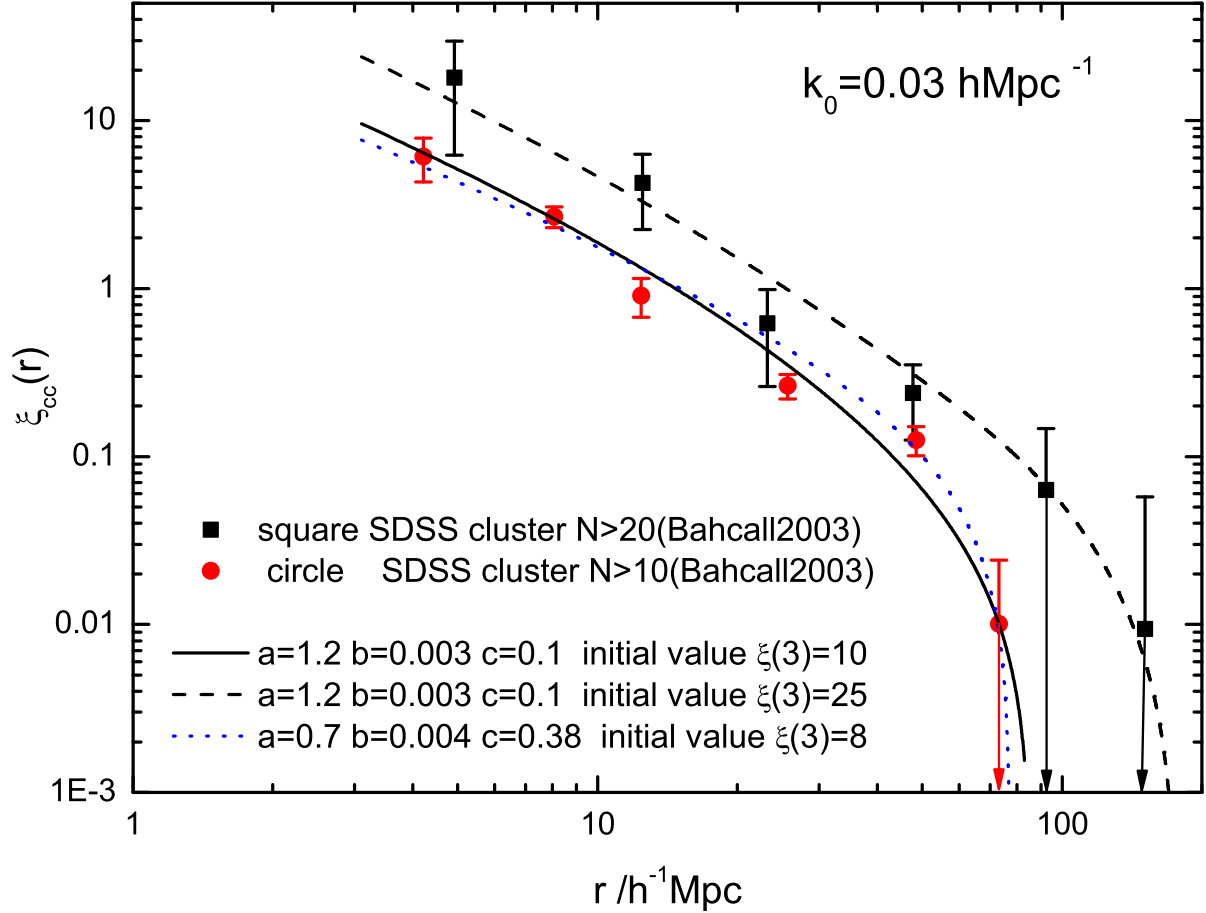}
\caption{The solution $\xi_{cc}(r)$ for clusters
confronts the data of SDSS clusters with tow types of richness \cite{Bachall03}.
 Notice that here  $(a,b,c)$ are the same as for galaxies,
however,
$k_0=0.03$ hMpc$^{-1}$ is taken for clusters,
 smaller than that for galaxies.}
\label{BahcallFig}
\end{figure}
In Figure \ref{BahcallFig},
for each set $(a,b,c)$ ,
two solutions $\xi_{cc}(r)$ with different amplitudes are given
to compare with two sets of data with richness $N> 10$ and  $N> 20$
from the SDSS  \cite{Bachall03}.
To match the data of clusters of $N> 20$,
we have chosen a greater boundary condition $\xi(r_b)$ than that
of $N> 10$,
while $k_0$ is the same.
This results in a higher correlation amplitude
and an apparently longer ``correlation length''
for the $N> 20$  clusters.
Interpreted by the field equation, Eq.(\ref{eqfinal}),
the $N> 20$  clusters have a greater $m$
than the $N> 10$ clusters.
The solutions match the data  available
on the whole range $r=(4\sim 100)$h$^{-1}$Mpc,
and there is no small-scale insufficiency of correlation
that occurred for the galaxy case.
This indicates that, to account for the correlation of clusters,
the order of $(\delta\psi)^2$ is accurate enough
in the perturbation treatment of our formulation.
Since $k_0=0.03$ Mpc$^{-1}$ for clusters and $k_0=0.055$ Mpc$^{-1}$ for galaxies,
it can be inferred that
the mean density $\rho_0$ involved in this cluster survey
should be lower by $(0.03/0.055)^2\sim 0.3$
than those in the galaxy case.

\begin{figure}
\centering
\includegraphics[width=\linewidth]{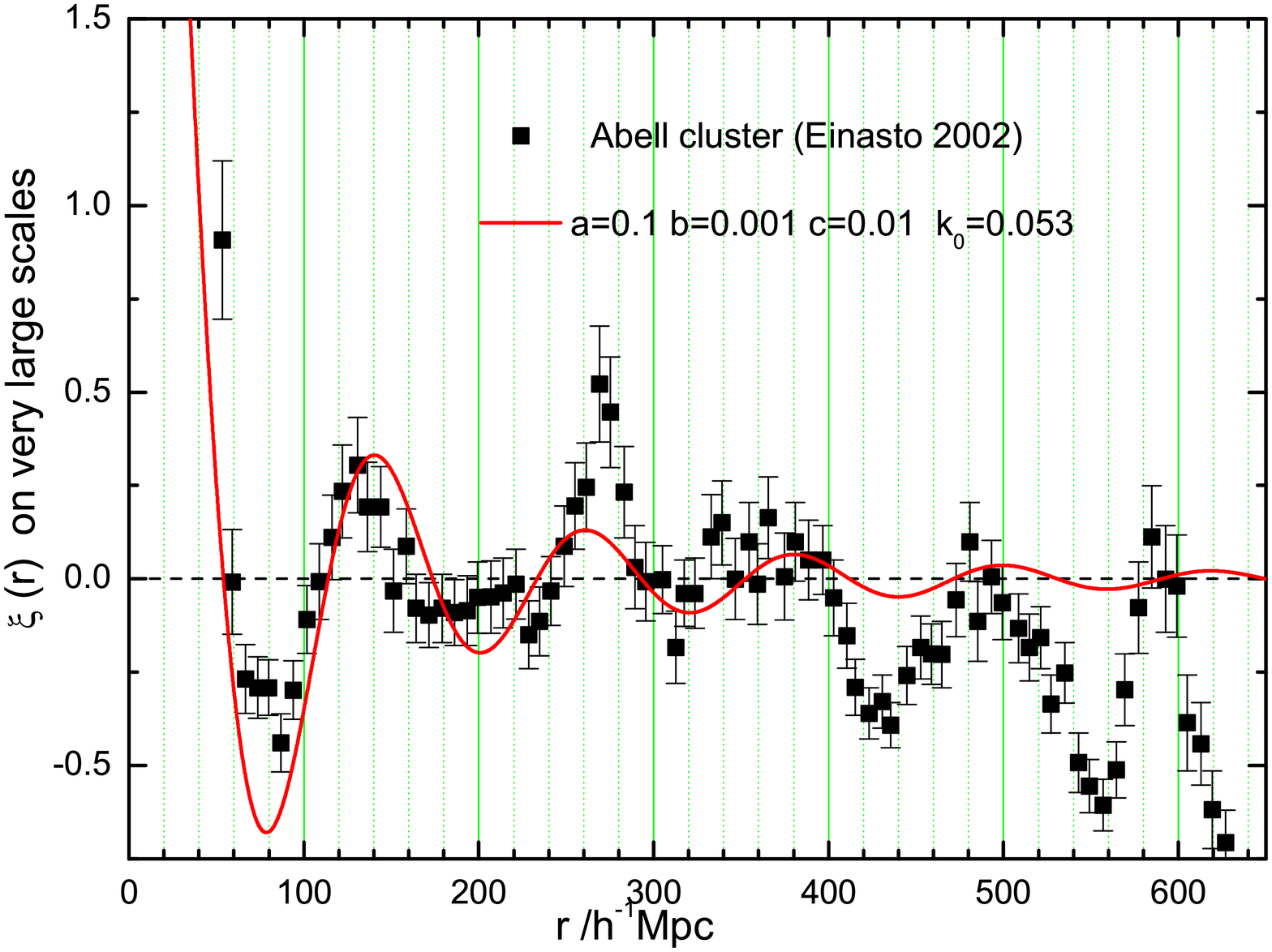}
\caption{The solution $\xi(r)$ with a small $a$
has periodic oscillations on very large scales.
$a=0.1$ is taken for demonstration.
The characteristic length is $\lambda_0  \sim 120$ h$^{-1}$Mpc.
It qualitatively explains the data of
Abell X-ray clusters \cite{Einasto02}.}
\label{oscillation}
\end{figure}

\begin{figure}
\centering
\includegraphics[width=\linewidth]{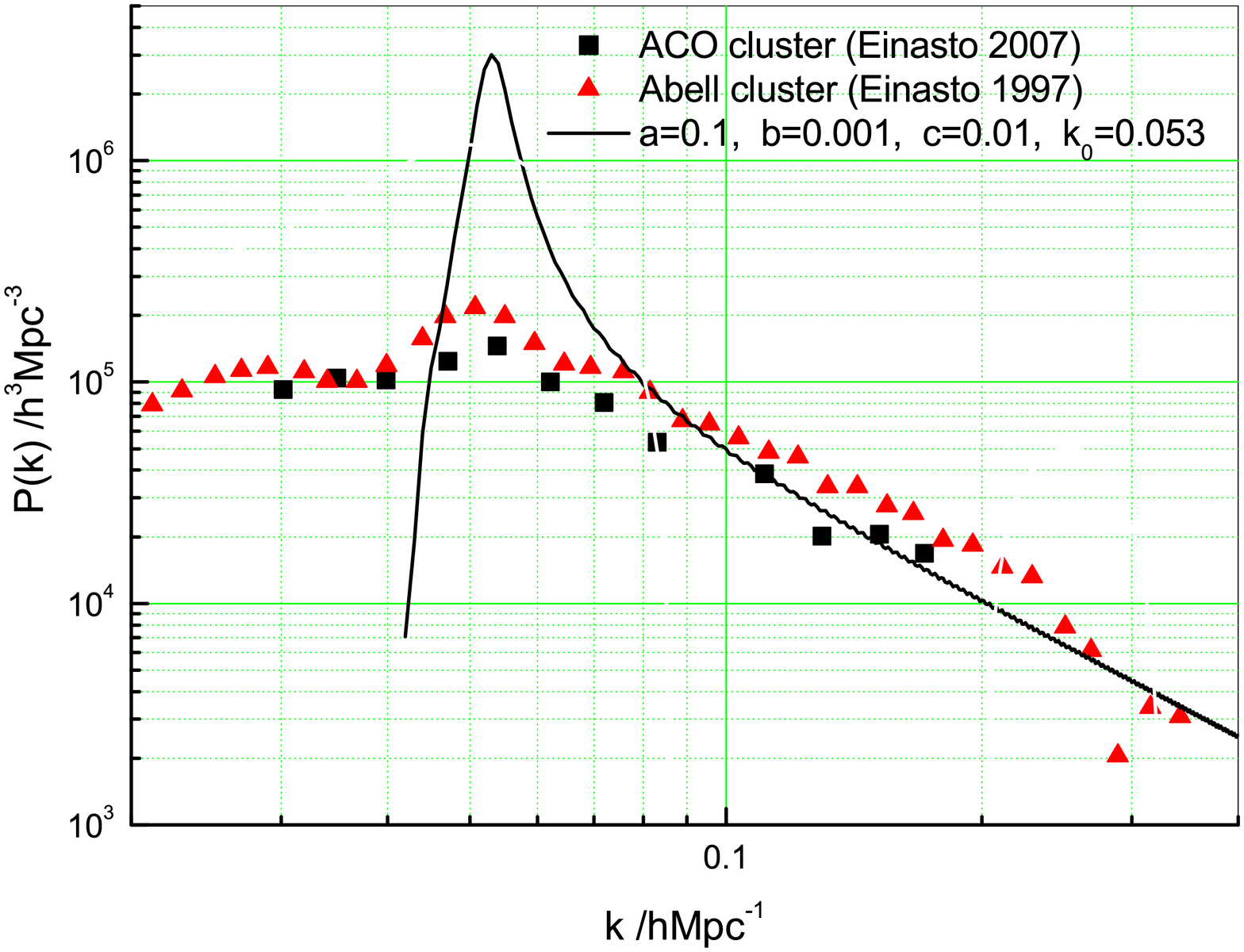}
\caption{ The converted $P(k)$ has a prominent peak at $k=k_0$,
corresponding to the periodic oscillations
of $\xi(r)$ in Fig \ref{oscillation}.
The profile of $P(k)$ qualitatively agrees with the observational data
of clusters \cite{Einasto2000b,EinastoNature}.
}
 \label{ClusterPk}
\end{figure}

It has long been known that,
there is a scaling behavior,
that is,
the cluster correlation scale increases
with the mean spatial separation  between clusters
\cite{SzalaySchramm,BahcallWest1992,Bahcall1999,Croft, Gonzalez}.
For a power-law $\xi_{cc} =(r_0/r)^{1.8}$ fitting,
the data indicates a ``correlation length"
\be  \label{scaling}
r_0 \simeq 0.4d_i,
\ee
where $d_i=n_i^{-1/3}$ and $n_i$
is the mean number density of clusters of type $i$.
For SDSS, the scaling can be also fitted by
$r_0 \simeq 2.6 d_i\,^{1/2} $  \cite{Bachall03},
and for the 2df galaxy groups $r_0\simeq 4.7 d_i\,^{0.32}$\cite{Zandivarez}.
From these surveys,
the common pattern is that $r_0$ increases with $d_i$.
This kind of  $r_0-d_i$ scaling has been a theoretical challenge \cite{Bahcall1999},
and was thought to be caused
by a fractal distribution of galaxies and clusters \cite{SzalaySchramm}.
In our theory  the scaling is fully embodied in
the solution $\xi_{cc}(k_0 r)$
with the characteristic wavenumber $k_0=(8\pi G m n/c_s^2)^{1/2} \propto d^{-3/2}$.
To comply with the empirical power-law,
we take the theoretical ``correlation length" as
$r_0(d) \propto \xi_{cc}^{1/1.8}$,
where $\xi_{cc}$ is the theoretical solution and depends on  $d$.
Fig.\ref{scaling} shows that
the solution $\xi_{cc}$ with $k_0=0.03$ hMpc$^{-1}$
gives the scaling  $r_0(d)\simeq 0.4d$,
agreeing well with the observation \cite{Bahcall1999}.
If  a greater $k_0=0.055$ hMpc$^{-1}$ is taken,
the solution $\xi_{cc}$ would yield a flatter scaling  $r_0(d)\simeq 0.3d$,
which seems to fit the data of APM clusters better\cite{Bachall03}.
This comparison tells that
a higher background density $\rho_0$
corresponds to a flatter slope of the scaling  $r_0(d)$.
Thus the  $r_0-d_i$ scaling is naturally interpreted
by the solution $\xi_{cc}(k_0r)$.
\begin{figure}
\includegraphics[width=\linewidth]{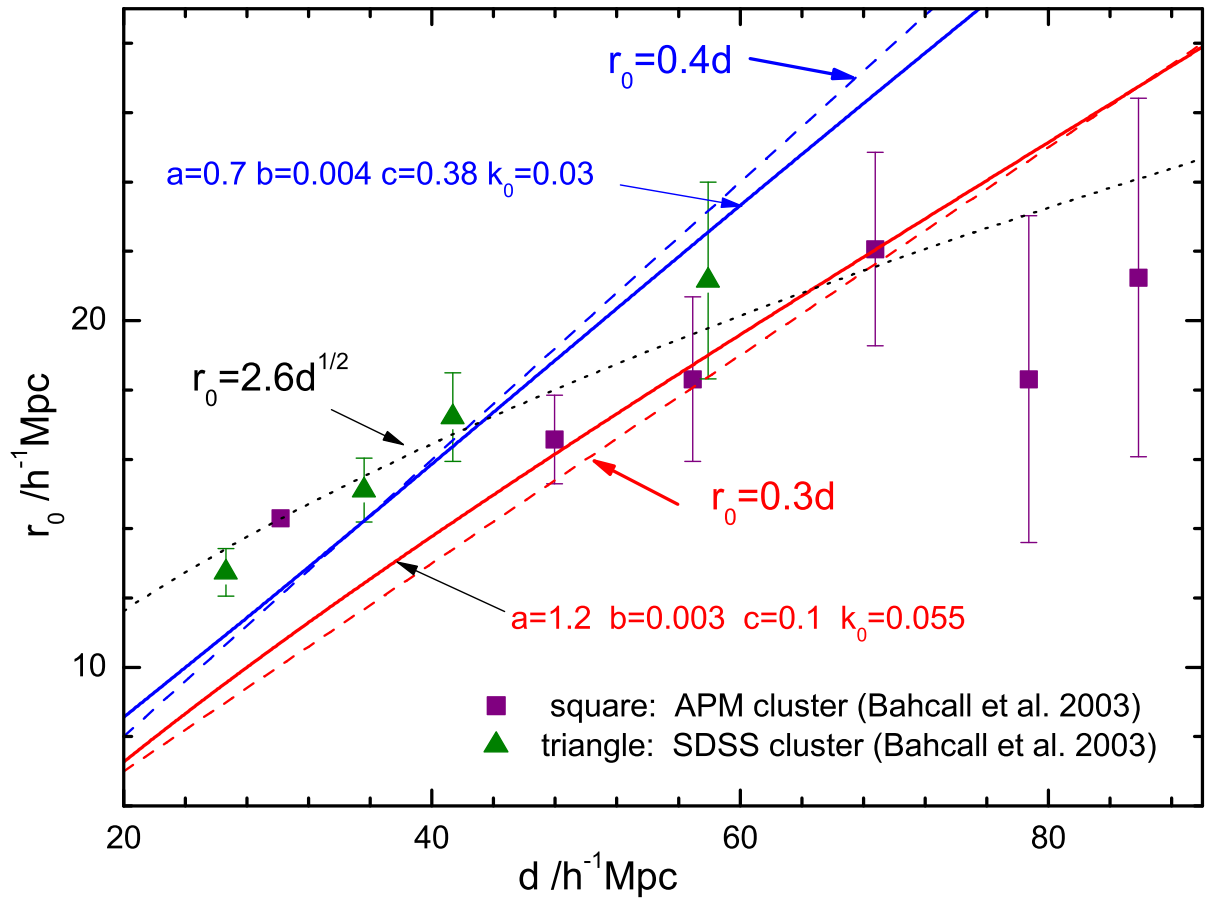}
\caption{The solution $\xi_{cc}(r)$ with $k_0=0.03$ hMpc$^{-1}$
gives the scaling  $r_0\simeq 0.4d$.
But with a greater $k_0=0.055$ hMpc$^{-1}$,
$\xi_{cc}$ would give a flatter scaling $r_0\simeq 0.3d$,
which seems to fit the data of APM clusters better \cite{Bachall03}.}
\label{scaling}
\end{figure}
Extended to very large scales,
the observed $\xi_{cc}(r)$ exhibits
a pattern of periodic oscillations
with a characteristic wavelength $\sim 120$Mpc\cite{Einasto1997a,EinastoNature}.
This  behavior was originally
found in the galaxy distribution in narrow pencil beam surveys \cite{Broadhurst},
also occurred in the correlation function of galaxies \cite{Tucker},
and of quasars \cite{Yahata}.
There have been also various interpretations on this periodic oscillations,
and one is that these correspond to the superclusters
of the comparable size \cite{Bahcall91}.
In Figure \ref{oscillation},
the theoretical $\xi(r)$ with small values $(a,b,c)$
exhibits periodic oscillations,
which is close to the Gaussian solution \cite{zhang}.
To achieve the characteristic wavelength $\lambda_0 = 2\pi/k_0 \sim 120$Mpc,
one needs $k_0\simeq 0.053$ Mpc$^{-1}$.
To yield high oscillations, a small  $a=0.1$ is taken for demonstration.
The data of the Abell X-ray clusters is also plotted \cite{Einasto02},
exhibiting the prominent, periodic oscillations.
Qualitatively,
the solution $\xi(r)$ agrees with
the pattern of oscillation of the data,
but has a damped amplitude at increasing $r$.
The power spectrum $P(k)$ converted from the solution $\xi(r)$ with $a=0.1$
does have a prominent peak,
as in Fig. \ref{ClusterPk}  \cite{EinastoNature,Einasto2000b}.
Thus in our theory
this kind of oscillations
originates from the field equation itself with a  sufficiently small viscosity.

\section{ Conclusions and Discussions }

We have presented a field theory of density fluctuations of
a Newtonian gravitating system,
applied it to the study of the correlation functions of galaxies and
of clusters in a homogeneous, isotropic Universe.

As the key setup,
we have obtained the field equation (\ref{masseq}) of the mass density field $\psi$,
under the condition of thermal equilibrium  or hydrostatic equilibrium.
It suits the studying of the mass distribution of  Universe.
This approach is different from those using the gravitational potential.
In dealing with the high nonlinearity,
we have written the field as $\psi=\psi_0+\delta\psi$,
the order $(\delta\psi)^2$ has been kept in  perturbations.
The generating functional $Z[J]$ of the correlation functions
has been written down as an path integral over $\psi$.
The field equation (\ref{eqfinal}) of $\xi=\langle\delta\psi\delta\psi\rangle$
has been derived as the main result,
whereby the Kirkwood-Groth-Peebles ansatz
and renormalization have been used.
The equation is Helmholtz-like and nonlinear,
with three parameters $(a,b,c)$ representing the nonlinear effects
beyond the Gaussian approximation.
Notably,
the characteristic wavelength $\lambda_0$ occurs as the only scale,
and the mass $m$ appears in the source.
By the dependence on $m$ and  $\lambda_0$,
the equation simultaneously explains
several longstanding, seemingly unrelated
features of the clustering,
such as
the profile similarity of $\xi_{cc}$ of clusters
to $\xi_{gg}$  of galaxies,
the differences in amplitude and in correlation length
of $\xi_{cc}$ and $\xi_{gg}$,
the scaling behavior $r_0 \simeq 0.4d$,
and the pattern of periodic oscillations in $\xi_{cc}$
with a wavelength $\lambda_0 \sim 120$Mpc.

The  solution $\xi_{gg}$ for fixed  $(a,b,c)$
agrees with
the observational data of the galaxy surveys
over a range $r=(1\sim 50)$Mpc.
So do the associated power spectrum,
projected correlation, and angular correlation.
With the same set of $(a,b,c)$,
but with a greater $m$ and a longer $\lambda_0$,
the solution  $\xi_{cc}$ also matches the data of clusters
over a range $r=(4\sim 100)$h$^{-1}$Mpc.
Thus, our theory sheds light on the understanding of
the clustering and the large scale structure of Universe.

There are several issues and possible extensions of the current theory.

1,  As is seen,  the amplitude of theoretical $\xi_{gg}$ at $r \leq 1$ Mpc
is lower than the observational data of galaxies.
This may indicate that the actual clustering of galaxies
requires higher order terms of the fluctuation
beyond $(\delta\psi)^2$.
To include  $(\delta\psi)^3$ and the higher,
the treatment  will become more involved
and the occurrence of  $G^{(4)}$, in addition to $G^{(3)}$,
will be anticipated in the field equation of $\xi_{gg}$.
This extension will be our future work.

2, The formulation established in this paper can be systematically used
to derive the field equations of $G^{(3)}$, etc,
which will be inevitably more complicated.

3, In this paper we have not considered the influence of the cosmic dark energy,
nor a possible bias of clustering by baryon.
These would need more refined studies.

4,  Finally,  in this paper
the effect of the expansion of the Universe
has not been considered.
Thus, it would be desired that
an extension could be made to the case of the cosmic evolution.

\section*{Acknowledgements}
Y. Zhang's research work has been supported by
the CNSF No.11073018, 11275187, SRFDP, and CAS.

\appendix
\numberwithin{equation}{section}

\section[]{ Stratonovich-Hubbard Transformation
       and  Grand Partition Function as a Path Integral}

From the identity
\begin{align}
&\exp{ \left[ \frac{1}{2}m^2V \right]}
=\frac{1}{ \sqrt{2\pi V}}
\int_{-\infty}^{\infty}dx
\exp{ \left[ -\frac{1}{2V} x^2 +m x  \right]}, \nonumber\\
&\,\,\, {\rm with}\,\,V>0,
\end{align}
one can extend to the Stratonovich-Hubbard identity \cite{Stratonovich,Hubbard}:
\begin{align}
&\exp{ \left[ \frac{1}{2}\sum_{i,j}^N  m_i V_{ij} m_j \right]}
=\frac{1}{ \sqrt{\det(2\pi V)}}\nonumber\\
&\cdot\prod_i^N \int_{-\infty}^{\infty}dx_i
\exp{ \left[ -\frac{1}{2}\sum_{i,j}^N  x_iV^{-1} _{ij} x_j
                   +\sum_i ^N  x_i m_i  \right]},
\end{align}
where $(V_{ij})$ is a symmetric matrix with positive eigenvalues.
This can be further extended to the continuous case.
Let $V( r)$ be a long range attractive potential, and its inverse
$K$  as a kernel is defined  by
\be
\int d^3r   K( {\bf r}_1-{\bf r} )  V({\bf r}-{\bf r}_2)
=\delta^{(3)}({\bf r}_1- {\bf r}_2).
\ee
Then the Stratonovich-Hubbard identity in this continuous case is \cite{Zinn-Justin}
\begin{align}
&\exp{ \left[ \frac{1}{2}T^{-1} \sum_{i,j} V(r_i-r_j) \right]}
=
\mathcal{N} \int_{-\infty}^{\infty}D \phi\nonumber\\
&\exp{ \left[- \frac{1}{2} T
\int d^3r_1 d^3r_2 \phi(r_1)K(r_1-r_2) \phi(r_2)
                   +\sum_i ^N \phi(r_i)  \right]},
\end{align}
where the numerical factor $\mathcal{N}\propto 1/\sqrt{ \det V}$
is a multiplicative factor to the grand partition function $Z$,
irrelevant to the ensemble averages of physical quantities,
thus can be dropped.
We mention that,  for a formally stricter treatment,
a hard core of radius $r_c$, say the size of a typical galaxy,
should have been introduced at the center of $V(r)$
so that there would be a cutoff of lower limit of integration to avoid
the divergence.
But this divergence will only occur in $\mathcal{N}$
and is dropped off eventually.

The interesting case is the  potential
$V({\bf r}_1- {\bf r}_2) = \frac{1}{|{\bf r}_1- {\bf r}_2|}$.
By
\be
\nabla^2 \frac{1}{|{\bf r}_1- {\bf r}_2|}
           =-4\pi \delta^{(3)}({\bf r}_1- {\bf r}_2),
\ee
the kernel is
$ K( {\bf r}_1-{\bf r} )
= - \frac{1}{4\pi}\delta^{(3)}( {\bf r}_1-{\bf r})\nabla^{2}$.
Integrating by parts yields:
\be
\int d^3r_1 d^3r_2 \phi(r_1)K(r_1-r_2) \phi(r_2)
  = \frac{1}{4\pi}\int d^3r (\nabla \phi)^2,
\ee
so that
\begin{align}
&\exp{ [ \frac{1}{2} T^{-1} \sum_{i,j}^N
\frac{Gm^{2} }{|{\bf r}_i- {\bf r}_j|} ]}
=\nonumber\\
&\int_{-\infty}^{\infty}D \phi
\exp{ [- \frac{1}{2}  \alpha
     \int d^{3}r (\nabla \phi)^2
+\sum_i ^N \phi(\bf r_i) ]},
\label{partialintegral}
\end{align}
where $\alpha \equiv T/4\pi  Gm^2 $.
The term $\sum_i \phi(\bf r_i)$ in Eq.(\ref{partialintegral})
is a sum of interactions of the field $\phi$
with the $i$-point mass at $\bf r_i$
(and could also be written as an integration
$\sum_i \phi({\bf r_i})= \int d^3r \phi({\bf r})n(\bf r)$
where $n(\bf r)$ is the number density of particles).

We use the above result to write
the grand partition function  $Z$ in Eq.(\ref{Z1}) as a path integral.
The kinetic energy term in $e^{-H/T}$ after integrating over
the momentum $ d^3p_i$ gives
\be
\int \frac{d^3p_i }{(2\pi)^3}
                       \exp[-p_i^2/2mT]=(\frac{mT}{2\pi})^{3/2}.
\ee
The potential term in $e^{-H/T}$ is given by Eq.(\ref{partialintegral}),
in which only $\sum_i ^N \phi(r_i)$ involves
integration over the coordinate $d^3r_i$  and gives
\be
\int \prod_{i=1}^N d^3r_i \exp{ \sum_i ^N \phi(r_i)}=
\left[\int d^3r  \exp{  \phi(r)} \right]^N.
\ee
Thus one has
\begin{align}
 \label{ZA}
&Z=\sum_{N=0}^\infty \frac{1}{N!}[z(\frac{mT}{2\pi })^{3/2} ]^N\nonumber\\
&\cdot\int_{-\infty}^{\infty}D \phi
   \exp{ \left[- \frac{1}{2}
  \alpha \int d^3r (\nabla \phi)^2 \right]}
\left[\int d^3r  \exp{  \phi(r)} \right]^N  \nonumber \\
&=\int_{-\infty}^{\infty}D \phi
\exp{ \left[- \frac{1}{2}
  \alpha \int d^3r (\nabla \phi)^2
  + z(\frac{m T }{2\pi})^{3/2}\int d^3r e^{\phi(r)} \right]}
\end{align}
Using the fugacity $z=(2\pi/mT)^{3/2}n_0$
for a dilute gas of the mean number density $n_0$,
one finally obtains
\be
Z=\int D \phi \,
   e^{-\alpha\int d^3 r  [\frac{1}{2 }(\nabla \phi)^2-k_J^2e^{\phi}]},
\ee
with $k_{J}^{2}\equiv 4\pi G\rho_{0}/c_{s}^{2}$,  $\rho_0=m n_0$ and $c^2_s=T/m$.
This is  Eq.(\ref{Z2}) in the context.

\section{Derivation of  Field Equation
                      and  Renormalization}

We present the derivation of the field equation of
the 2-pt correlation function $G^{(2)}(\textbf{r}-\textbf{r}^{\prime})$.
The technique involved is the functional differentiation of
the generating functional $Z[J]$ in Eq.(\ref{ZJ})
with respect to the external source $J$.
The method is commonly adopted in field theory of particle physics and
of condensed matter physics\cite{Goldenfeld}.
We start with the ensemble average of Eq.(\ref{field_eq})
of the mass density field in the presence of  $J$,
\be \label{main}
\langle\nabla^{2}\psi({\bf r} )-\frac{1}{\psi({\bf r} )}(\nabla\psi({\bf r} ))^{2}
          +k_{J}^{2}\psi({\bf r} )^{2}+J\psi({\bf r} )^{2}\rangle_J  =0,
\ee
differentiate it with respect to $J$
\begin{equation} \label{main}
\frac{\delta}{\delta J({\bf r} ^{\prime})}
\langle\nabla^{2}\psi({\bf r} )-\frac{1}{\psi({\bf r} )}(\nabla\psi({\bf r} ))^{2}
+k_{J}^{2}\psi({\bf r} )^{2}+J({\bf r})\psi({\bf r} )^{2}\rangle_J=0,
\end{equation}
and set $J=0$, and will end up with
the field equation for $G^{(2)}({\bf r}-{\bf r}')$.
 In the following we deal with each term of Eq.(\ref{main}).

The first term of Eq.(\ref{main}) has
\be \label{1b2}
\langle\nabla^{2}\psi({\bf r})\rangle _J
= \nabla^{2}\langle\psi( {\bf r} )\rangle_J .
\ee
Changing the ordering of $\frac{\delta}{\alpha \delta J({\bf r} ^{\prime})} $
and $\nabla^{2}$,
using the definition
\be \label{Gdef}
G^{(2)}({\bf r}-{\bf r}')
=\frac{\delta}{\alpha \delta J({\bf r}')} \langle\psi({\bf r})\rangle_J|_{J=0},
\ee
we obtain
\be \label{nablaG}
\nabla^{2}\left(\frac{\delta}{\alpha \delta J(\bf{r}^{\prime})}
\langle\psi({\bf r})\rangle_J\right)|_{J=0}
= \nabla^{2}G^{(2)}(\bf{r}-\bf{r}^{\prime}).
\ee
For the remaining three terms of Eq.(\ref{main}),
we firstly work in  the Gaussian approximation  \cite{zhang}, i.e,
at the lowest order of fluctuation,
\be
\langle\frac{(\nabla\psi)^{2}}{\psi}\rangle_J
 \simeq \frac{ (\nabla \langle\psi\rangle_J)^{2}}{\langle\psi\rangle_J},
\ee
$\langle\psi(\textbf{r})\rangle|_{J=0} =\psi_{0}=1$,
$\nabla \psi_0=0$, so that the second term vanishes.
The third and fourth terms involve
\be
\langle \psi^2 \rangle_J \simeq \langle\psi\rangle_J^2.
\ee
By  Eq.(\ref{Gdef}) and
\be \label{deltaJ}
\frac{\delta J({\bf r})}{\delta J({\bf r}')}
=\delta^{(3)}({\bf r}-{\bf r}^{\prime}),
\ee
Eq.(\ref{main}) in the Gaussian approximation reduces to
the Helmholtz equation with a point source
\be \label{Helmoltz}
\nabla^{2}G^{(2)}({\bf r} )+k_{0}^{2}G^{(2)}({\bf r} )
          =-\frac{1}{\alpha } \delta^{(3)}({\bf r} ),
\ee
where $k_0=\sqrt{2}k_J$ is the characteristic wavenumber.
The term $+k_0^{2}G^{(2)}$
has a plus sign because  gravity is attractive.
The Gaussian solution  is of the form:
\be \label{GGaussian}
G^{(2)}({\bf r}) \propto \frac{Gm}{c^2_s}\frac{\cos(k_0 r)}{r},\,\,\,\,\,\,
\frac{Gm}{c^2_s}\frac{ \sin (k_0 \, r)}{r},
\ee
subject to certain boundary condition in specific applications.
From Eq.(\ref{GGaussian}), we find an important property
that the amplitude is proportional to the mass $m$:
\be \label{proptom}
G^{(2)} \propto m.
\ee
By Fourier transform
\be
G^{(2)}({\bf r})=\int P(k)e^{i\bf k\cdot r}d^3r
\ee
the power spectrum in the Gaussian approximation is
\be \label{Pkgaussian}
P(k)=\frac{1}{2n_0}\frac{1}{(k/k_0)^2-1},
\ee
telling that $P(k)$ is higher for galactic
objects with a lower spatial number density $n_0$.
We just mention that
the similar form of  Eq.(\ref{Helmoltz})
also occurs in the the Gaussian approximation of the Landau-Ginziburg theory
of phase transition \cite{Goldenfeld,Binney},
where $G^{(2)}(\bf r)$ is also called the bare propagator.
However, in Landau-Ginziburg theory,
the corresponding term $-\mu^2G^{(2)}(\bf r)$, in place of $+k_0^22G^{(2)}(\bf r)$,
 has a negative sign,
the solution is $G^{(2)}({\bf r}) \propto e^{-\mu r}/r$
with $1/\mu$ being the correlation length.
At the critical point of phase transition, $\mu\rightarrow 0$,
$G^{(2)}({\bf r}) \propto 1/r$,
the correlation becomes  long-range.
In contrast, in our case,
 gravity is a long range attractive interaction,
the self-gravitating system is long-range correlated,
as evidenced by
the fact that  $G^{(2)}({\bf r}) $ in Eq.(\ref{GGaussian})
has $\cos(k_0 r)$ and $\sin(k_0 r)$,
instead of the exponential decay $e^{-\mu r}$.
In this sense, the self-gravitating system is always
at the critical point of phase transition \cite{Saslaw}.

Now beyond the Gaussian approximation,
we shall include high order terms of the fluctuation $\delta \psi$,
in calculation of in the remaining three terms of Eq.(\ref{main}).

The third term of Eq.(\ref{main}) is simple and has
\begin{align}
\langle \psi^{2}\rangle_J
=\langle(\langle\psi \rangle+\delta\psi)^{2}\rangle_J
=\langle\psi \rangle_J^{2}+\langle\delta\psi\delta\psi\rangle_J,
\end{align}
where $\langle \delta \psi \rangle_J=0$ is used.
Applying $\frac{\delta}{\alpha \delta J(\bf{r}')}$ to the above yields
\begin{align} \label{3b2}
&k_{J}^{2}\frac{\delta}{\alpha\delta J(\textbf{r}^{\prime})}
( \langle\psi \rangle_J^{2}+\langle\delta\psi\delta\psi\rangle_J
)_J|_{J=0}
=2\psi_{0}k_{J}^{2}  G^{(2)}(\textbf{r}-\textbf{r}^{\prime})\nonumber\\
 &+ k_{J}^{2}G^{(3)}(\textbf{r},\textbf{r},\textbf{r}^{\prime}),
\end{align}
where  the 3-pt correlation function
$ G^{(3)}({ \bf r},{\bf r},{ \bf r}')=\\
\frac{\delta}{\alpha \delta J( {\bf r}')}\langle\delta\psi\delta\psi\rangle|_{J=0}$
is used.
In our previous treatment  \cite{zhang2009nonlinear},
$G^{(3)}$ in the above was dropped as a high-order term.

The fourth term of Eq.(\ref{main}) is also simple and has
\be
\langle J\psi^{2}\rangle_J= J \langle \psi^{2}\rangle_J
=J\langle\psi \rangle_J ^{2}+J\langle\delta\psi\delta\psi\rangle_J,
\ee
and, by Eq.(\ref{deltaJ}),
it gives
\be \label{4b2}
\frac{\delta}{\alpha \delta J(\textbf{r}^{\prime})}\langle J\psi^{2}\rangle_J |_{J=0}
=\frac{1}{\alpha}[\psi_{0}^{2}+G^{(2)}(0)]\delta^{(3)}(\bf{r}-\bf{r}^{\prime}).
\ee
Here $G^{(2)}(0)=\langle\delta\psi\delta\psi\rangle
           =\lim_{r'\rightarrow r}G^{(2)}({\bf r}-{\bf r}')$.
This quantity might be divergent as ${\bf r}'\rightarrow \bf r$.
Of course, for the system of galaxies,
the definition of $G^{(2)}(\bf r)$ applies
only for $r> r_c$, the galaxy size.
The occurrence of the quantity $G^{(2)}(0)$,
and later also $\nabla G^{(2)}(0)$ and $\nabla^2 G^{(2)}(0)$,
is inevitable
when high order terms of $\delta\psi$  are included beyond the Gaussian approximation.
This is common in calculating the 2-point correlation function
in any field theory with interactions,
both in particle physics and condensed matter physics.
In the former case, the analogue of $G^{(2)}(0)$ is divergent,
and, in the latter, a cutoff is introduced for $|{\bf r}-{\bf r}'|\ge r_c$,
and $G^{(2)}(0)$ is finite.
For example, in our case,
it could be expressed as an integration over the momentum
\be
G^{(2)}(0)=\lim_{r\rightarrow r'}
            \int d^3k \frac{e^{-i{\bf k\cdot({\bf r}-{\bf r}') }}}{k^2-k_0^2}
=\int d^3k \frac{1}{k^2-k_0^2}
\ee
of the ``bare" propagator $1/(k^2-k_0^2)$ of the Gaussian approximation.
$\nabla G^{(2)}(0)$ and $\nabla^2 G^{(2)}(0)$
will have the similar expressions, correspondingly.
These three quantities at zero separation $r=0$ are undetermined
in our case for the system of galaxies.
In fact, as in field theory,
one is usually not interested in the specific values of these quantities
at all.
The standard procedure to handle these quantities
is the well-known renormalization.
These quantities are eventually absorbed into  the physical quantities,
such as  the mass, the field, the coupling constant, etc,
depending on the specific field theory concerned \cite{Binney}.
In this paper, similarly,
we shall also  adopt the practice of renormalization to
 absorb $ G^{(2)}(0)$,  $\nabla G^{(2)}(0)$, and $\nabla^2 G^{(2)}(0)$.

The second term of Eq.(\ref{main}) is more involved,
as it has a factor $\frac{1}{\psi}$.
To deal with this term systematically,
we expand it in terms of the fluctuation  $\delta\psi$ up to
\be \label{expand}
\frac{1}{\psi}=\frac{1}{ \langle  \psi \rangle  +\delta\psi}
\simeq
\frac{1}{ \langle  \psi \rangle }
\left(1-\frac{\delta\psi}{ \langle  \psi \rangle }+
(\frac{\delta\psi}{\langle  \psi \rangle })^{2} \right).
\ee
(We skip the subscript ``$J$" in $\langle  \psi \rangle_J$
 temporarily in the following for simple notation.)
As an approximation,
this perturbation is accurate  only for $\delta\psi/\langle\psi\rangle \ll 1$.
At small scales,
 $\delta\psi$ can be larger than $\langle\psi\rangle$,
so it can be anticipated that
the small-scale high nonlinearities
may not be sufficiently accounted for in this order of perturbations.
By Eq.(\ref{expand}), one has
\begin{align}\label{2}
\langle\frac{(\nabla\psi)^{2}}{\psi}\rangle
&=\frac{1}{ \langle \psi \rangle}\langle\left(1-\frac{\delta\psi}{\langle\psi \rangle}
  +(\frac{\delta\psi}{\langle  \psi \rangle })^{2} \right)
(\nabla  \langle\psi\rangle +\nabla\delta\psi)^{2} \rangle\nonumber\\
&\simeq \frac{(\nabla\langle\psi\rangle)^{2}}{\langle\psi\rangle}
+\frac{\langle(\nabla\delta\psi)^{2}\rangle}{\langle\psi\rangle}
-\frac{2\nabla \langle  \psi \rangle}{ \langle  \psi \rangle^{2}}\cdot \langle\delta\psi\nabla\delta\psi\rangle\nonumber\\
&+\frac{(\nabla \langle  \psi \rangle )^{2}}{ \langle  \psi \rangle^{3}}\langle(\delta\psi)^{2}\rangle,
\end{align}
where $\langle\delta \psi\rangle=0$ is used,
and $(\delta\psi)^3$ and higher have been dropped.
(\ref{2}) contains four sub-terms.
The first and second terms of (\ref{2})
will be treated together.
In our previous treatment \cite{zhang2009nonlinear},
$\langle(\nabla\delta \psi)^{2}\rangle\rightarrow\nabla^{2}\langle(\delta\psi)^{2}\rangle$
was taken, which was not correct.
Now we treat it in the following.
By the field equation (\ref{field_eq}), one has
\be\label{(1)}
 (\nabla\psi)^{2}=\psi\nabla^{2}\psi+(k_{J}^{2}+J)\psi^{3},
\ee
and there is an identity:
\be
(\nabla\psi)^{2}=\nabla\cdot(\psi\nabla\psi)-\psi\nabla^{2}\psi\label{(2)}
=\frac{1}{2}\nabla^2 \psi^2  -\psi\nabla^{2}\psi.
\ee
Eq.(\ref{(1)}) and Eq.(\ref{(2)}) are added  together yield
\begin{align}  \label{(3)}
(\nabla\psi)^{2}=\frac{1}{4}\nabla^{2}\psi^{2}+\frac{1}{2}(k_{J}^{2}+J)\psi^{3}.
\end{align}
Taking the ensemble average of Eq.(\ref{(3)}),
we have
\be
\langle(\nabla\psi)^{2}\rangle=\frac{1}{4}\langle\nabla^{2}\psi^{2}\rangle
+\frac{1}{2}(k_{J}^{2}+J)\langle\psi^{3}\rangle
\ee
Substituting $\psi=\langle\psi\rangle+\delta\psi$ into both sides leads to
\begin{align}
&(\nabla \langle\psi\rangle)^{2}+\langle(\nabla\delta\psi)^{2}\rangle
=\frac{1}{4}\nabla^{2}\langle\psi \rangle^{2}+\frac{1}{4}\nabla^{2}\langle\delta\psi\delta\psi\rangle
\nonumber\\
&+\frac{1}{2}(k_{J}^{2}+J) \langle  \psi \rangle^{3}
+\frac{3}{2}(k_{J}^{2}+J) \langle  \psi \rangle\langle\delta\psi\delta\psi\rangle,
\end{align}
where $\langle\delta\psi \rangle=0$ is used, and
the higher order term $\langle(\delta\psi)^{3}\rangle$ is dropped.
Thus, the first and second sub-terms of (\ref{2})  together
give
\begin{align} \label{sub12}
&\frac{(\nabla  \langle  \psi \rangle)^{2}}{ \langle  \psi \rangle}
+\frac{\langle(\nabla\delta\psi)^{2}\rangle}{ \langle  \psi \rangle}
=
\frac{1}{4} \frac{\nabla^{2}\langle  \psi \rangle^{2}}{\langle  \psi \rangle}
+\frac{1}{4}\frac{\nabla^{2}\langle\delta\psi\delta\psi\rangle}{\langle  \psi \rangle}\nonumber\\
&+\frac{1}{2}(k_{J}^{2}+J)\langle  \psi \rangle ^{2}
+\frac{3}{2}(k_{J}^{2}+J) \langle\delta\psi\delta\psi\rangle.
\end{align}
Applying  $\frac{\delta}{\delta J({\bf r} ^{\prime})} $ to Eq.(\ref{sub12})
and setting $J=0$ yields the contribution of
the first and second terms of Eq.(\ref{2}):
\begin{align}  \label{2.a+2.b}
&\frac{\delta}{\alpha  \delta J(\textbf{r}^{\prime})}
[\frac{(\nabla  \langle  \psi \rangle)^{2}}{ \langle  \psi \rangle}
+\frac{\langle(\nabla\delta\psi)^{2}\rangle}{ \langle  \psi \rangle} ]|_{J=0}\nonumber\\
&= \frac{1}{2}\nabla^{2} G^{(2)}(\textbf{r}-\textbf{r}^{\prime})+k_{J}^{2}\psi_{0}G^{(2)}(\textbf{r}-\textbf{r}^{\prime})\nonumber\\
&-\frac{1 }{4\psi_{0}^{2}} \nabla^{2}G^{(2)}(0)\cdot G^{(2)}(\textbf{r}-\textbf{r}^{\prime}) \nonumber\\
&+(\frac{1 }{4\psi_{0}}\nabla^{2}+\frac{3}{2} k_{J}^{2} )G^{(3)}(\textbf{r},\textbf{r},\textbf{r}^{\prime})\nonumber\\
&+\frac{1}{2\alpha }[\psi_{0}^{2}+3G^{(2)}(0)]\delta^{(3)}(\textbf{r}-\textbf{r}^{\prime}).
\end{align}
where
$\nabla^{2} G^{(2)}(0)=\nabla^{2}\langle\delta\psi\delta\psi\rangle
=\lim_{r\rightarrow 0}\nabla^{2}G^{(2)}({\bf r} )$.
The third term of (\ref{2}) yields
\begin{align}\label{b3}
&-2\frac{\delta}{\alpha \delta J(\textbf{r}^{\prime})}[\frac{\nabla\langle  \psi \rangle}{\langle  \psi \rangle^{2}}\cdot
\langle(\nabla\delta\psi)\delta\psi\rangle]|_{J=0}\nonumber\\
&=-\frac{\delta}{\alpha \delta J(\textbf{r}^{\prime})}[\frac{\nabla\langle  \psi \rangle}{\langle  \psi \rangle^{2}}\cdot
\langle\nabla(\delta\psi)^{2}\rangle]|_{J=0}\nonumber\\
&=-\frac{1}{\psi_{0}^{2}}\nabla G^{(2)}(0)\cdot\nabla G^{(2)}(\textbf{r}-\textbf{r}^{\prime}).
\end{align}
The fourth term of (\ref{2}) yields
\be \label{b4}
\frac{\delta}{\delta J(\textbf{r}^{\prime})}
[\frac{(\nabla \langle  \psi \rangle)^{2}}{\langle  \psi \rangle ^{3}}
      \langle(\delta\psi)^{2}\rangle]|_{J=0} =0
\ee
by $\nabla \psi_0=0$.
The sum of Eqs.(\ref{2.a+2.b}), (\ref{b3}) and (\ref{b4}) gives
the contribution of the second term of (\ref{main})
\begin{align} \label{2b2}
&-\frac{\delta}{\alpha \delta J(\textbf{r}^{\prime})}\langle\frac{(\nabla\psi)^{2}}{\psi}\rangle|_{J=0}\nonumber\\
&= -[\frac{1}{2}\nabla^{2}+k_{J}^{2}\psi_{0}
    -\frac{\nabla^{2}G^{(2)}(0)}{4\psi_{0}^{2}}]G^{(2)}(\textbf{r}-\textbf{r}^{\prime})\nonumber\\
    &\resizebox{.9\hsize}{!}{$+\frac{1}{\psi_{0}^{2}}\nabla G^{(2)}(0)\cdot\nabla G^{(2)}(\textbf{r}-\textbf{r}^{\prime})
-(\frac{1}{4\psi_{0}}\nabla^{2}
   +\frac{3}{2}k_{J}^{2})G^{(3)}(\textbf{r},\textbf{r},\textbf{r}^{\prime})$}\nonumber\\
& -\frac{1}{2\alpha }[\psi_{0}^{2}+3G^{(2)}(0)]\delta^{(3)}(\textbf{r}-\textbf{r}^{\prime}).
\end{align}

Now,
plugging  Eqs.(\ref{nablaG}), (\ref{3b2}), (\ref{4b2}), and (\ref{2b2})
into Eq.(\ref{main}),
we  obtain the equation of 2-pt correlation function:
\begin{align} \label{25}
&(\nabla^{2}+2k_{J}^{2}\psi_{0}) G^{(2)}(\textbf{r}-\textbf{r}^{\prime}) \nonumber\\
&+[\frac{1}{2\psi_{0}^{2}}\nabla^{2}G^{(2)}(0) G^{(2)}(\textbf{r}-\textbf{r}^{\prime})
-(\frac{1}{2\psi_{0}}\nabla^{2}+k_{J}^{2})G^{(3)}(\textbf{r},\textbf{r},\textbf{r}^{\prime})\nonumber\\
&+\frac{2}{\psi_{0}^{2}}\nabla G^{(2)}(0)\cdot\nabla G^{(2)}(\textbf{r}-\textbf{r}^{\prime})]\nonumber\\
&=-\frac{1}{\alpha }[\psi_{0}^{2}-G^{(2)}(0)]\delta^{(3)}(\textbf{r}-\textbf{r}^{\prime}).
\end{align}
This is just Eq.(\ref{2pt3pt}) in the text.

Observe that  this equation  is not closed for  $G^{(2)}$,
but contains   $G^{(3)}$, as is expected.
One would go on to get the field equation of  $G^{(3)}$, etc.
This kind of hierarchy is common to the field equation of correlations
in a nonlinear theory when the perturbation method is used.
To cut off the hierarchy and get a closed equation for  $G^{(2)}$,
we adopt the Kirkwood-Groth-Peebles ansatz
in  Eq.(\ref{Ansatz}) \cite{Kirkwood,groth1977large}.
For simplicity, taking $\textbf{r}^{\prime}$ be the origin $0$,
the ansatz is
\be
\resizebox{.9\hsize}{!}{$G^{(3)}(\textbf{r},\textbf{r},\textbf{r}^{\prime})
=G^{(3)}(\textbf{r},\textbf{r},0)
=Q[2G^{(2)}(0)G^{(2)}(\textbf{r})+(G^{(2)}(\textbf{r}))^{2}].$}
\ee
We mention that $G^{(3)}$ is of order $(\delta \psi)^3$
and $G^{(2)}$ is of order $(\delta \psi)^2$,
therefore,
the use of ansatz causes an increase of order of the terms containing $Q$ in perturbation.
Then one has
\begin{align}
&-(\frac{1}{2\psi_{0}}\nabla^{2}+k_{J}^{2})G^{(3)}(\textbf{r},\textbf{r},0)\nonumber\\
&=-Q[\frac{1}{\psi_{0}}G^{(2)}(0)\nabla^{2}G^{(2)}(\textbf{r})
+\frac{1}{\psi_{0}} (\nabla G^{(2)}(\textbf{r}))^{2}\nonumber\\
&+\frac{1}{\psi_{0}}G^{(2)}(\textbf{r})\nabla^{2}G^{(2)}(\textbf{r})
+2k_{J}^{2}G^{(2)}(0)G^{(2)}(\textbf{r})
+k_{J}^{2}(G^{(2)}(\textbf{r}))^{2}\nonumber\\
&+\frac{1}{\psi_{0}}G^{(2)}(\textbf{r})\nabla^{2}G^{(2)}(0)
+\frac{2}{\psi_{0}}\nabla G^{(2)}(0)\cdot \nabla G^{(2)}(\textbf{r})
].
\label{3in2}
\end{align}
Substituting  (\ref{3in2}) into (\ref{25}), we get
the field equation of $G^{(2)}(\textbf{r})$
\begin{align}
&(1-\frac{Q}{\psi_{0}}G^{(2)}(0))\nabla^{2}G^{(2)}(\textbf{r})\nonumber\\
&\resizebox{.8\hsize}{!}{$+[\frac{1}{2\psi_{0}^{2}}(1-2Q\psi_{0})\nabla^{2}G^{(2)}(0)
+2k_{J}^{2}\psi_{0}(1-\frac{Q}{\psi_{0}}G^{(2)}(0))]G^{(2)}(\textbf{r})$}\nonumber\\
&-Qk_{J}^{2}(G^{(2)}(\textbf{r}))^{2}
-\frac{Q}{\psi_{0}}G^{(2)}(\textbf{r})\nabla^{2}G^{(2)}(\textbf{r})
-\frac{Q}{\psi_{0}} (\nabla G^{(2)}(\textbf{r}))^{2}\nonumber\\
&+\frac{2}{\psi_{0}^{2}}(1-Q\psi_{0})\nabla G^{(2)}(0)\cdot \nabla G^{(2)}(\textbf{r})\nonumber\\
&=-\frac{1}{\alpha }[\psi_{0}^{2}-G^{(2)}(0)]\delta^{(3)}(\textbf{r}),
\label{G2}
\end{align}
which is closed in terms of $G^{(2)}(\textbf{r})$.
Now we introduce the notations
\begin{align}
&\textbf{a}\equiv \frac{2}{\psi_{0}^{2}}(1-Q\psi_{0})\nabla G^{(2)}(0),\\
&b\equiv \frac{Q}{\psi_{0}},\label{bdef}\\
&c\equiv \frac{Qk_{J}^{2}}{k_{0}^{2}},\label{cdef} \\
&k_{0}^{2}\equiv 2k_{J}^{2}\psi_{0}(1-bG^{(2)}(0))
 +\frac{1}{2\psi_{0}^{2}}(1-2Q\psi_{0})\nabla^{2}G^{(2)}(0).\label{k_0}
\end{align}
Then  (\ref{G2}) becomes
\begin{align} \label{coeff2}
&\resizebox{.9\hsize}{!}{$[1-b G^{(2)}(0)-bG^{(2)}(\textbf{r})] \nabla^{2}G^{(2)}(\textbf{r})
+k_{0}^{2}(1 -cG^{(2)}(\textbf{r}))G^{(2)}(\textbf{r})$} \nonumber\\
&+[{\bf a}-b\nabla G^{(2)}({\bf r})] \cdot\nabla G^{(2)}(\textbf{r})
    =-\frac{1}{\alpha }[\psi_{0}^{2}-G^{(2)}(0)]\delta^{(3)}(\textbf{r}).
\end{align}
Note that the parameter $Q$
 has been absorbed into $\bf a$, $b$, $c$,
 and the latter will be regarded as independent parameters.
Let us do renormalization.
The first term on l.h.s of (\ref{coeff2}) has
$Z_0  \equiv 1-b G^{(2)}(0)$
as part of the coefficient of $\nabla^{2}G^{(2)}(\textbf{r})$,
which can be absorbed in the definition of $G^{(2)} (\bf r)$.
Since $G^{(2)} \propto (\delta\psi)^2$,
this amounts to the renormalization of the density field $\delta \psi$.
Explicitly, multiplying Eq.(\ref{coeff2}) by $Z_0^{-2}$
and making the following substitutions
\begin{align}
& G^{(2)}({\bf r}) \rightarrow G^{(2)}_R({\bf r}) \equiv Z_0^{-1} G^{(2)}({\bf r}),
 \hspace{.5cm} k_0^2 \rightarrow k_{0R}^2 \equiv Z_0^{-1} k_0^2, \nonumber \\
& c \rightarrow c_R \equiv Z_0 c,
\hspace{.5cm} {\bf a} \rightarrow {\bf a}_R \equiv Z_0^{-1} {\bf a},\\
 &\psi_0^2
        \rightarrow \psi_{0R}^2 \equiv Z_0^{-2} (\psi_0^2-G^{(2)}(0)) ,
\end{align}
 Eq.(\ref{coeff2}) finally becomes
\begin{align} \label{renormeq}
&(1-bG^{(2)}(\textbf{r}))\nabla^{2}G^{(2)}(\textbf{r})
+k_{0}^{2}(1- cG^{(2)}(\textbf{r}))G^{(2)}(\textbf{r}) \nonumber\\
&+({\bf a}-b\nabla G^{(2)}({\bf r})) \cdot\nabla G^{(2)}(\textbf{r})
    =-\frac{1}{\alpha }\psi_0^2 \delta^{(3)}(\textbf{r}),
\end{align}
where all the quantities ${\bf a}$, $c$, $k_0^2$, $G^{(2)}(\bf r)$, and $\psi_0^2$
are all understood to be the renormzlized ones,
their subscript ``$R$" being dropped for simple notation.
The renormalized characteristic wavenumber
$k_{0}$ is given by $k^2_0=  2 k_{JR}^{2} =8\pi G m_R^2n_0/T$,
where
\be
m_R^2\equiv  m^2\psi_0 + \frac{T}{8\pi G n_0}
             Z_0^{-1}\frac{1}{2\psi_{0}^{2}}(1-2Q\psi_{0})\nabla^{2}G^{(2)}(0).
\ee
$m_R$ is the renormalized mass in place of the ``bare"  mass $m$,
and its subscript ``$R$" will also be dropped from now on for simple notation.
Thus, by the renormalization  procedure,
$\nabla^{2}G^{(2)}(0)$ has been absorbed by $m$,
$\nabla    G^{(2)}(0)$ absorbed by $\bf a$,
and $G^{(2)}(0)$ by $\psi_0$, $\delta\psi$, $Q$, and others.
After the renormalization,
 one can set  $\psi_0=1$
in Eq.(\ref{renormeq}).

When ${\bf a}=0$, $b=0$, $c=0$,
Eq.(\ref{renormeq}) reduces to the Helmholtz equation in Eq.(\ref{Helmoltz})
in the Gaussian approximation.
Thus, the terms involving ${\bf a}$, $b$ and $c$
are the nonlinear effects at the order $(\delta\psi)^2$
beyond the Gaussian approximation.
By isotropy of the system,
it is simpler to write the field equation (\ref{renormeq}) in the radial direction.
Denoting
$\xi(x) \equiv G^{(2)}(\textbf{r})$, $x\equiv k_{0}r$,
and
$\xi'\equiv \frac{d}{dx}\xi(x)$,
then Eq.(\ref{renormeq}) becomes
\begin{align} \label{equation}
\resizebox{.9\hsize}{!}{$(1-b\xi)\xi''+
( (1-b\xi)\frac{2}{x}+a )\xi' +\xi  -b \xi'\, ^{2} -c\xi^{2}=
-\frac{1}{\alpha }    \frac{\delta(x)k_0}{x^{2}}$}
\end{align}
where $a\equiv |{\bf a}/k_0|$.
The dimensionless parameters $a$, $b$, and $ c$ will be regarded
as independent,
even though they essentially
come from the combinations of
the quantities $Q$, $ G^{(2)}(0)$, $\nabla G^{(2)}(0)$, and $\nabla^2G^{(2)}(0)$.
The parameter $a$ in Eq.(\ref{equation}) plays a role of the effective viscosity.
The nonlinear terms  $\xi'^2$ and $\xi^2$
can enhance the correlation at small scales.
Compared with Eq.(8) in Reference \cite{zhang2009nonlinear},
now Eq.(\ref{equation}) has a new term $ -b\xi$
in the coefficients of $ \xi''$  and  $\frac{2}{x}$,
and a new term $-c\xi ^2$.
As we have checked,
for the values  $a$, $b$, $ c$  taken in confronting the observational data of surveys,
the numerical solutions of the two equations
differ only slightly.

\end{document}